\documentclass[lettersize,journal]{IEEEtran}
\usepackage{amsmath,amsfonts}
\usepackage{algorithmic}
\usepackage{algorithm}
\usepackage{array}
\usepackage[caption=false,font=normalsize,labelfont=sf,textfont=sf]{subfig}
\usepackage{bm}
\usepackage{textcomp}
\usepackage{stfloats}
\usepackage{url}
\usepackage{verbatim}
\usepackage{graphicx}
\usepackage{cite}
\begin{document}

\title{Network Topology Optimization via Deep Reinforcement Learning}

\author{Zhuoran Li,
        Xing Wang,
        Ling Pan,
        Lin Zhu,
        Zhendong Wang,
        Junlan Feng,~\IEEEmembership{Fellow,~IEEE,}
        Chao Deng,
        Longbo Huang,~\IEEEmembership{Senior Member,~IEEE,}
\thanks{Zhuoran Li, Ling Pan and Longbo Huang are with IIIS, Tsinghua University, Beijing, China. Xing Wang, Lin Zhu, Zhendong Wang, Junlan Feng and Chao Deng are with China Mobile Research Institute, Beijing, China.}
\thanks{Corresponding author: Longbo Huang (longbohuang@tsinghua.edu.cn)}}

\markboth{IEEE TRANSACTIONS ON COMMUNICATIONS, JANUARY 2023}%
{Shell \MakeLowercase{\textit{et al.}}: A Sample Article Using IEEEtran.cls for IEEE Journals}


\maketitle

\begin{abstract}
Topology impacts important network performance metrics, including link utilization, throughput and latency, and is of central importance to network operators. However, due to the combinatorial nature of network topology, it is extremely difficult to obtain an optimal solution, especially since topology planning in networks also often comes with management-specific constraints. As a result, local optimization with hand-tuned heuristic methods from human experts is often adopted in practice. Yet, heuristic methods cannot  cover the global topology design space while taking into account constraints, and cannot guarantee to find good solutions. 

In this paper, we propose a novel deep reinforcement learning (DRL) algorithm for graph searching, called DRL-GS, for network topology optimization. DRL-GS consists of three novel components, including a verifier to validate the correctness of a generated network topology, a graph neural network (GNN) to efficiently approximate  topology rating, and a DRL agent to conduct a topology search. {DRL-GS can efficiently search over relatively large topology space and output topology with satisfactory performance.} We conduct a case study based on a real-world network scenario, and our experimental results demonstrate the superior performance of DRL-GS in terms of both efficiency and performance.
\end{abstract}


\vskip -0.2in
\section{Introduction}
\IEEEPARstart{D}{ue} to the rapid development of communication technologies, more and more devices are connected to the Internet. As a result, the network scale keeps increasing and the network infrastructure is continuously upgraded, e.g., by adopting large-capacity fiber optic cables, to keep up with the need for better service quality. 
Since main network performance metrics, such as link utilization, throughput and latency, are heavily affected by the network structure, 
network operators pay much attention on topology optimization as a critical problem.

However, the network planning problem is  very challenging. First, the topology problem is combinatorial in nature. Thus, it is often of high complexity, i.e., exponential in the number of links. Second, to conduct network topology optimization, there are technical issues caused by complex management-specific  constraints (often nonlinear and even nonconvex) in network topology optimization, due to operation requirements,  e.g., in terms of the allowed fraction of changed links, overall modification costs, or network performance such as link utilization after optimization. 
Therefore, different models and approaches have been proposed for various network topology planning scenarios. For instance, \cite{bondarenko2019optimization,haile2020data,gerstel2014multi}  formulate the network planning problem as a mixed integer linear programming problem or a complex multi-objective optimization problem focusing on cost minimization and multi-layer recovery. 
Existing topology optimization works have developed different algorithms to solve the network topology problem with different formulations, such as minimum spanning tree algorithm \cite{li2005design}, centralized connection algorithm \cite{ramanathan2000topology} and approximation algorithm\cite{williamson2011design}.
\cite{halim2019combinatorial,rezoug2018guided,santos2011optimization} develop heuristic methods, which focus on explicit objective functions in limited connections and are unable to extend in larger networks.
Yet, these algorithms are not guaranteed to achieve close-to-optimal performance, and  can only be applied to specified scenarios. Moreover, existing methods typically incur high computational complexity, primarily due to the combinatorial nature of network topology planning. 
Therefore, to support the rapidly increasing demand of high network service quality, there is an urgent need to develop efficient topology optimization methods to support network capacity expansion and improve operating performance through efficient searching.

In this work, we propose a novel solution to solve the network topology optimization problem utilizing deep reinforcement learning (DRL) for graph searching, which is called DRL-GS. 
DRL has been demonstrated to achieve superior performance in many scenarios, e.g., designing molecular structures \cite{you2018graph}, sharing bike scheduling \cite{pan2019deep} and wireless communication \cite{he2021overview}. 
DRL-GS builds on the generalization power of DRL and consists of three key novel components, namely, a topology verifier to validate the correctness of a generated network topology, a graph neural network (GNN)  to efficiently approximate the topology rating, and an RL agent to conduct topology search. The formal procedure of DRL-GS is shown in Fig. \ref{fig:DRL_framework} below. 

DRL-based algorithms are proposed for solving topology optimization, e.g. Pointer Network\cite{vinyals2015pointer} and NeuroPlan\cite{zhu2021network}, and more general combinatorial optimization problems, 
e.g., learning cuts for integer programming \cite{tang2020reinforcement}, vehicle routing problem \cite{nazari2018reinforcement}, online computing offload \cite{huang2019deep},  and TSP problem \cite{joshi2020learning}. 
Our work distinguishes itself from these results in the following.  First, DRL-GS adopts a GNN for learning complex objective functions of the combinatorial problem. This enables us to efficiently rate the goodness of topologies  
to facilitate searching.  Second, DRL-GS introduces a novel action space compression to avoid searching over an overly-large space (exponential in network size with dozens of nodes). Third, DRL-GS contains a topology verifier, which efficiently testifies the found solution and generates data for training GNN.

The main contributions of our paper are as follows. 

First, we formulate a  network topology optimization problem $\mathtt{NetTopoOpt}$, which takes into account topology adjustment feasibility, adjustment cost, and  performance impact. $\mathtt{NetTopoOpt}$ provides a general modeling framework for optimizing combinatorial network topology.

Second, we propose a general DRL-based scheme for graph searching named DRL-GS, which consists of three key novel components, a topology verifier  to validate the correctness of a generated network topology, 
a GNN to efficiently approximate the topology rating, and an RL agent to conduct topology search. {The novel design of DRL-GS allows us to efficiently search over a relatively large topology space. }

Third, we carry out a case study based on a real-world network topology optimization scenario from China Mobile \cite{chinamobiledataset} and conduct extensive experiments. {Our results demonstrate that the DRL-based algorithm for graph search outperforms existing heuristic search algorithms in both small action space, i.e., with  $O(2^{20})$ actions and $8$ nodes, and large search space, i.e., with  $O(10^{21})$ actions and $23$ nodes. }

\section{Problem definition}\label{chap:probdef}
In this section, we present the network topology optimization formulation.  Our overall objective is to find an optimal network topology, starting from a given initial structure, to optimize a given complex performance metric, subject to management-specific constraints on the adjustment. Below, we  present the general framework, based on which we present our solution framework. In the experiment section, we will carry out a detailed implementation based on a concrete setting. 
Specifically, for a given network, we denote its topology structure as  $G=(V,E)$, where $V=\{v_1, v_2, ..., v_N\}$ is the set of all vertices and $E=\{e=(v_m, v_n), v_m, v_n\in V\}$ is the set of all edges in the network. We define the adjacent matrix $\bm{x}$ of the topology $G$ as a variable of the connectivity property to optimize, i.e., if $i$ and $j$ are connected, $(\bm{x})_{ij}=1$, else $(\bm{x})_{ij}=0$. We will use the term topology and $\bm{x}$ interchangably in this paper.
We denote the adjacency matrix of the initial topology as $\bm{x}_0$. We sometimes write the set of edges $E$ as $E(\bm{x})$ to clearly indicate that the set of edges $E$ is a function of $\bm{x}$. 
Note that $\bm{x}$ can have $2^{\frac{N(N-1)}{2}}$ values in reality. 

In this work, we consider the following general constrained network topology  optimization problem, which we refer to as $\mathtt{NetTopoOpt}$ (see Section \ref{chap:casestudy} for a case study where we give concrete forms for the functions): 

\vskip -0.2in
\begin{eqnarray} 
(\mathtt{NetTopoOpt})\,\, \max: &f(\bm{x})=U(\bm{x}) + \gamma Cost(\bm{x}, \bm{x}_0), \label{eq:obj}\\
\text{s.t.}& Dist(e) \le D_{max}, \ \forall e\in E(\bm{x}), \label{eq:con-phy}\\
&Load(e)\le L_{max},\ \forall e\in E(\bm{x}), \label{eq:con-load}\\
& M(\bm{x}) = True.  \label{eq:con-feasibility}
\end{eqnarray}

{We now explain all the terms in the $\mathtt{NetTopoOpt}$ problem.} 
Here $U(\bm{x})$  denotes the 
performance of the network under the topology,  e.g., utilization of the network links. 
This term captures the performance aspect of the network topology for network operators to specify  service requirements. 
It is a nonlinear representation of the input and often involves complex computation in practical scenarios, {e.g., certain network performance metrics have nonlinear dependency on the input value.} 
$Cost(\bm{x}, \bm{x}_0)$ is the cost of transforming from ${\bm{x}_0}$ to the target value $\bm{x}$. 
This cost is usually proportional to the difference of the links in ${\bm{x}_0}$ and $\bm{x}$, due to deletion of old links and installation of new links.  
$\gamma<0$ is a weight between performance and cost.

We now explain the constraints. 
In the first constraint \eqref{eq:con-phy}, {$Dist(e)$ denotes the distance of two nodes connected by link $e$. Constraint \eqref{eq:con-phy} concerns the distance feasibility of the network topology adjustment, in that for each link $e=(v_i, v_j)$ in the new topology, any two connected nodes $v_i$ and $v_j$ need to be within connection distance $D_{max}$, e.g., due to fiber cable length in a wired network or due to the wireless radio connection range in a mobile network. }
Constraint \eqref{eq:con-load} is on the  utilization of the network links. Here $Load(e)$ denotes the utilization of the link, and $L_{max}$ is the maximum allowed utilization level. 
In practice, there are often constraints on link utilization, e.g., $90\%$ \cite{lin2012power}. 
Constraint \eqref{eq:con-feasibility} is an abstract feasibility requirement of the topology, which is often due to network management requirements and allows operators to impose policy-based restrictions on the resulting network topology (similar to BGP \cite{yi2016initialization}  on routing).
{For instance, there can be a length constraint on any path with many nodes to prevent a too long path. Or there may exist connection requirements among different nodes due to the utilization property of nodes. This constraint is introduced to allow  higher flexibility in modeling.}

\vskip -0.1in
\begin{figure}[ht]
 \begin{center}
 \includegraphics[width=0.4\textwidth]{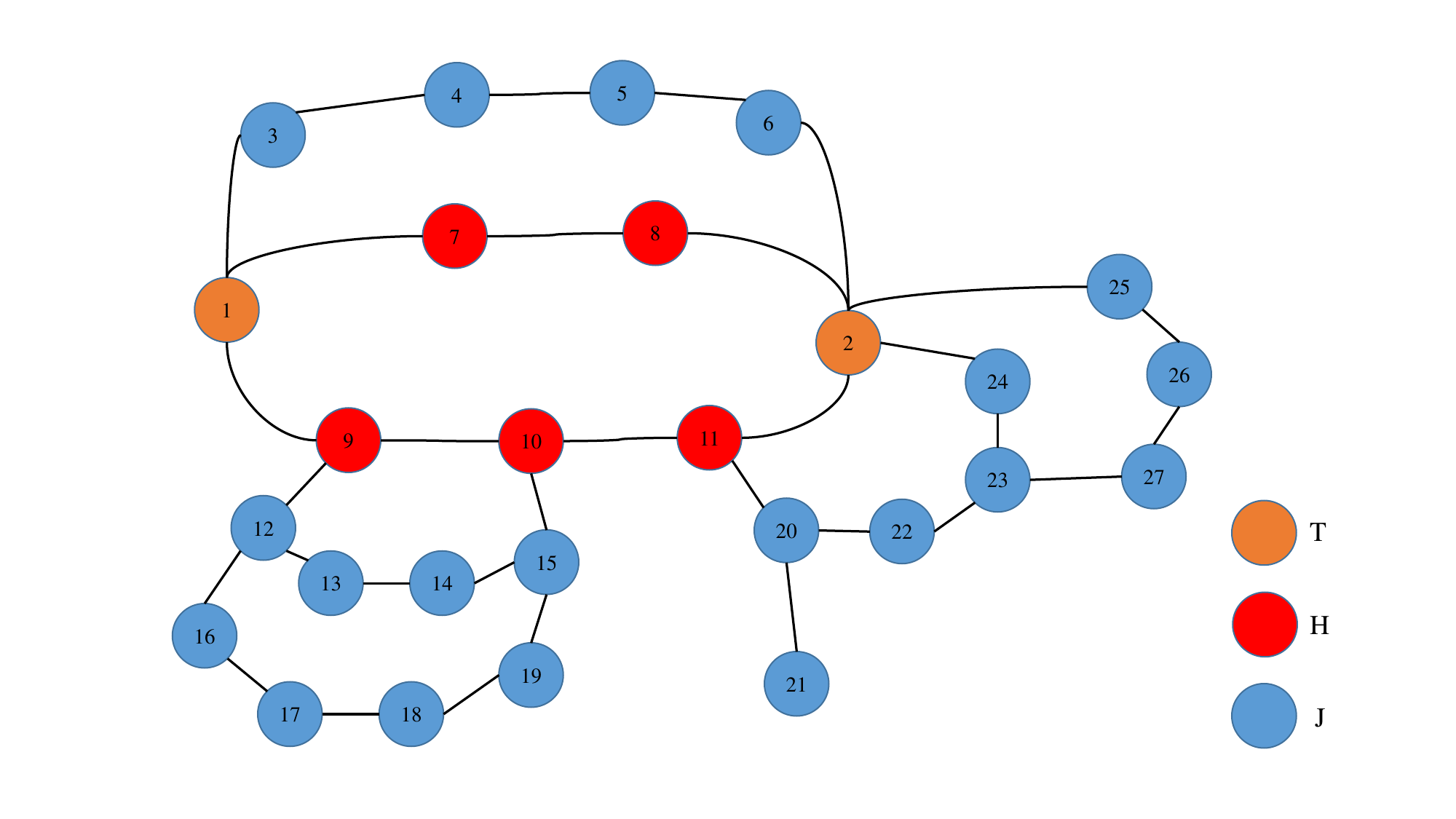}
 \vskip -0.2in
 \caption{{An example of a network topology. There are different node types and the formation of the network is constrained by the management requirements. }}
 \label{fig:example_network_planning}
 \end{center}
\end{figure}
\vskip -0.2in

Fig. \ref{fig:example_network_planning} gives a concrete example of a network topology. 
In this example, there are three different types of nodes, denoted as type-$T$ (orange), type-$H$ (red) and type-$J$ (blue), corresponding to nodes with different functionalities. In this example, the construction of such a network topology has many constraints. The distance constraint \eqref{eq:con-phy} in this example means that any edge connecting two nodes is bounded by a   maximum distance. 
The  constraint \eqref{eq:con-load} means that the aggregation of average traffic volume in type-$H$ nodes is no  larger than the maximum utilization of type-$T$ nodes. 
Take one requirement in Constraint \eqref{eq:con-feasibility} as an example, 
if a type-$J$ node and a type-$H$ node are connected, the maximum utilization, denoted by $U_{\max}(\cdot)$ (see Section \ref{chap:casestudy}),  of the type-$J$ node is no larger than that of the type-$H$ node for workload balance, e.g.,   $U_{\max}(v_{12})\le U_{\max}(v_{9})$.

{Note that our $\mathtt{NetTopoOpt}$ problem \eqref{eq:obj} is a general and abstract formulation, and offers a modeling framework for network topology optimization. }
However, from the definition and description of constraints, we see that the problem involves a complex combinatorial  objective function and nonlinear constraints. As a result, existing  algorithms, e.g., \cite{bondarenko2019optimization} and \cite{lou2011minimizing},  either cannot handle such complex functions or mostly lead to sub-optimal solutions and incur large computation complexity. Therefore, there is a strong need to design efficient algorithms. 

{Below, we describe our novel DRL-based method to handle general network topology optimization problems. Our method builds upon the generalization power of neural networks and introduces  techniques for handling the heavy computation requirement due to complex network structure. In the case study, we will show how the procedure can be  implemented when giving a specific problem.} 

\section{Methods and algorithms}\label{chap:methodsandalgorithms}
\begin{figure}[htp]
 \centering
 \includegraphics[width=0.4\textwidth]{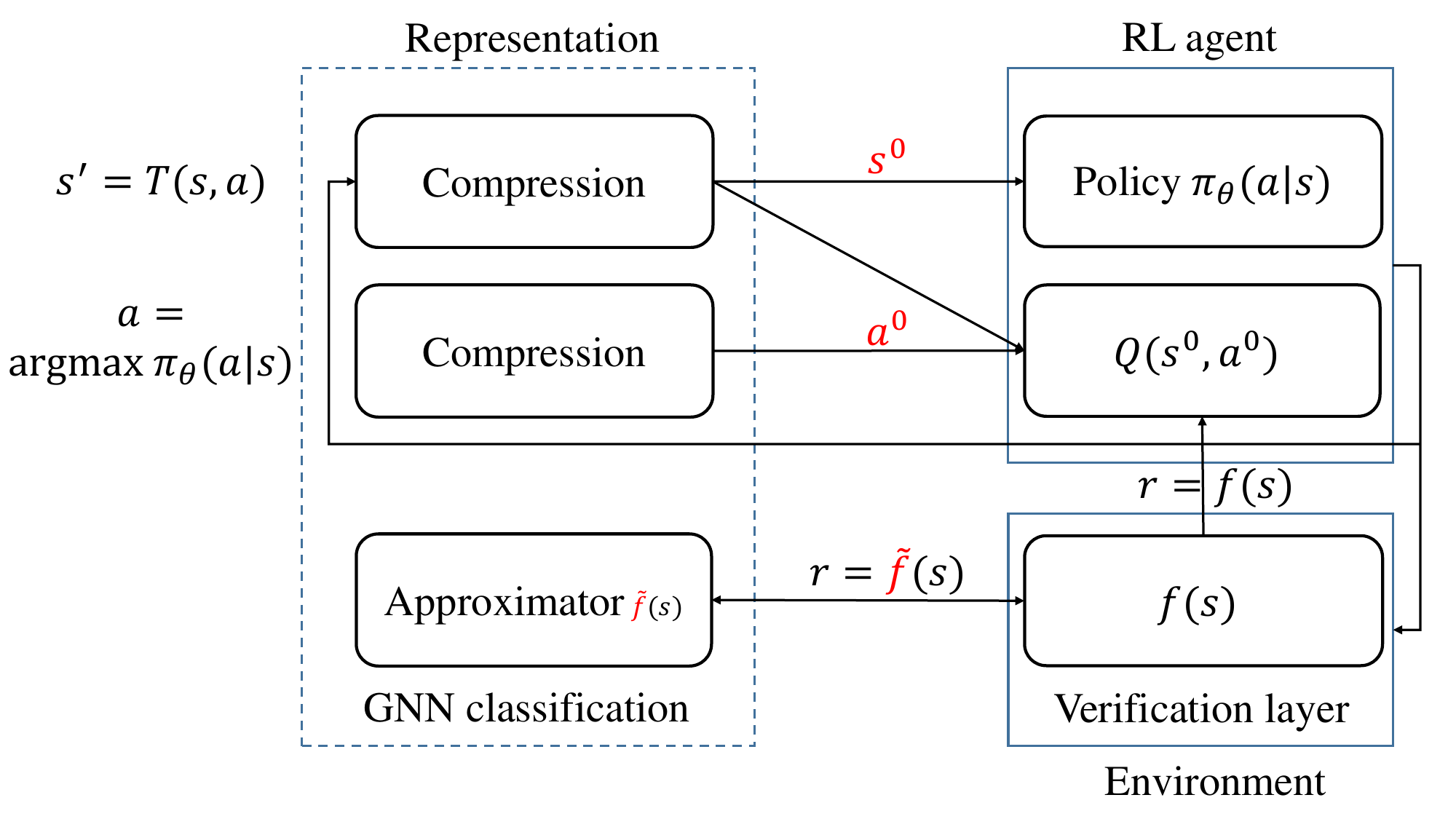}
 \vskip -0.2in
 \caption{The procedure of DRL-GS. There are three main components, i.e., the representation layer, the DRL agent and the verifier. Here $s$ denotes a state (topology), $a$ denotes an action (link change), $r$ denotes a reward function (score of topology). $T(s, a)$ denotes the transition function. $\pi_\theta$ denotes a parameterized policy. $f(s)$ is the objective value calculated by the verifier.  DRL-GS utilizes the representation layer to compress the state and action to $s_0$ and $a_0$ and the GNN classifier $\tilde{f}$ to learn the true objective function $f$.} 
 \label{fig:DRL_framework}
\end{figure}

{In this section, we present our DRL-based algorithm for graph searching, called DRL-GS. The  procedure of DRL-GS is shown in Fig. \ref{fig:DRL_framework} above. Specifically, DRL-GS contains a representation component, a DRL agent, and a topology verifier. }The representation component learns the network objective function and compresses the action and state spaces. It is introduced to reduce computation complexity due to large search space, i.e., $O(2^{\frac{N(N-1)}{2}})$ combinations, which makes even verifying the correctness of a topology difficult. 
{The DRL agent is used to learn control actions based on reinforcement learning updates  based on a DRL algorithm,  A2C or PPO \cite{mnih2016asynchronous,schulman2017proximal} in our framework, and works with the state, action and reward.}
The topology verifier is introduced to check feasibility of the output topology, so as to ensure with certainty that all constraints from network operators are guaranteed. It can also be used to help data generation in the large network setting (see Section \ref{chap:casestudy} for a concrete case study). 

\vskip -0.1in
\vskip -0.1in

{Below, we explain DRL-GS in detail.}
We denote the initial state space, {i.e., the set of all possible network topology,} as  $S=\{\bm{x}=(x_{ij}:i,j\in V): \bm{x} \,\,\text{feasible}\}$ (following the tradition, we will also use $s$ to denote a state when it is clear) and the action space, {i.e., the set of all possible connection pattern change,} as  $A=\{a=(a_{ij}:i,j\in V): a \,\,\text{feasible}\}$. Here each $a_{ij}\in\{0,1\}$ denotes whether an edge is present ($a_{ij}=1$) or not ($a_{ij}=0$). 
The next state is defined by the action and transition method $T(\cdot)$ that $x_{ij}(t+1)=T(x_{ij}(t),a_{ij} )= x_{ij}(t)+a_{ij}\ (\text{mod}\ 2)$. {In the algorithm, $\pi_\theta(a|s)$ denotes an update policy, which specifies a probability distribution over $a\in A$ given the current state $s$. The formal procedure of DRL-GS is shown in Algorithm \ref{alg:A2CGS}. }

\begin{algorithm}[htp]
\begin{algorithmic}[1]
\caption{{Deep reinforcement learning graph search algorithm (DRL-GS)}} %
\label{alg:A2CGS}
\STATE \textbf{Input}: Initial topology $G=(V,E)$, a GNN $\tilde{f}_{\phi(t)}$ with parameter $\phi(t)$, policy $\pi_\theta(a|s)$, replay buffer $\mathcal{B}$, timestep number $H$, epoch number $m$, threshold value $q$. \label{alga2cgs:inputstatement}
{\STATE Initialize GNN  $\tilde{f}_{\phi(1)}$, policy $\pi_\theta(a|s)$ and action choice $C_a$. Set reward choice $C_f=1$.  } \label{alga2cgs:initialization}
\FOR{epoch $i=1,...,m$}
\STATE Run DRL algorithm for $H$ steps with choice $C_f$ and $C_a$, compute the policy gradient and update the parameter $\theta$. If $C_f=1$, collect data $\{a^{j},s^{j},r^{j}\}_{j=1}^{H}$ into replay buffer $\mathcal{B}$. \label{alga2cgs:traina2c}
{
\STATE Update $\tilde{f}_{\phi(t+1)}\leftarrow\mathtt{GNNTraining}(\tilde{f}_{\phi(t)},\mathcal{B},q)$ (Algorithm \ref{alg:gcntrainingprocess}).  \label{alga2cgs:traingcn}
\STATE Generate data sequence $\mathcal{D}=\{s_j,a_j,r_j\}_{j=1}^n$ using policy $\pi_\theta$ and $\mathtt{Verifier}$ (Algorithm \ref{alg:generalverifier}) to validate and compute the objective values. Add the data to replay buffer $\mathcal{B}\leftarrow\mathcal{B}\cup\mathcal{D}$.\label{state:datagenerationfortest}
\STATE Run $C_f=\mathtt{GNNTesting}(\tilde{f}_{\phi(t)},\mathcal{D},q)$ (Algorithm \ref{alg:testinggcnprocess}). \label{alga2cgs:testinggcn}}
\ENDFOR
\end{algorithmic}
\end{algorithm}
\vskip -0.1in

DRL-GS uses $\mathtt{Verifier}$ (Algorithm \ref{alg:generalverifier}) to test if a topology is valid and  compute the objective value. {Note that this component is important, as it guarantees that the resulting solution is feasible.} 
In Section \ref{chap:casestudy}, we will show a concrete verifier algorithm based on specified network management requirements. 
{DRL-GS also provides RL agents for topology search, the $\mathtt{GNNTraining}/\mathtt{GNNTesting}$ algorithm for a classifier and action space choice for representation. }

\begin{algorithm}[htp]
\begin{algorithmic}[1]
\caption{General $\mathtt{Verifier}$ algorithm}
\label{alg:generalverifier}
\STATE \textbf{Input}: the initial topology $G(\bm{x_0})=(V,E(\bm{x_0}))$ with adjacent matrix $\bm{x_0}$, the target topology $G(\bm{x})=(V,E(\bm{x}))$. 
{
\STATE $\bm{I}_1=\cap_{e\in E}\mathbb{I}(Dist(e)\le D_{max})$ (Distance requirements), $\bm{I}_2=\cap_{e\in E}\mathbb{I}(Load(e)\le L_{max})$ (Utility requirements), $\bm{I}_3=\mathbb{I}(M(x)=True)$ (Feasibility requirements).
\IF {$\bm{I}_1=\bm{I}_2=\bm{I}_3=True$}
\STATE \textbf{Return} $f(\bm{x})=U(\bm{x}) + \gamma Cost(\bm{x}, \bm{x}_0)$. (Objective computation)
\ELSE
\STATE \textbf{Return} $f(\bm{x})=-\infty$.
\ENDIF
}
\end{algorithmic}
\end{algorithm}
\vskip -0.1in

{Step \ref{alga2cgs:initialization} initializes the value to choose the $\mathtt{Verifier}$ algorithm ($C_f=1$) or GNN ($C_f=2$) and the action space ($C_a=1$ means full space, $C_a=2$ means compressed space).} Choosing the compressed action space enables an efficient search over the state space. {Step \ref{alga2cgs:traina2c} trains the RL agent using the DRL algorithms, e.g., A2C or PPO \cite{mnih2016asynchronous,schulman2017proximal}.} During training, if we use the $\mathtt{Verifier}$ algorithm to directly calculate the objective value as the reward, i.e., $C_f=1$, we collect the data in the replay buffer for training GNN. Otherwise, i.e., $C_f=2$, we use $0$ or $1$ (label poor or good topology) as the reward. 
This is done to further reduce the complexity in providing detail rating, which can still face high computational complexity. 
%
Step \ref{alga2cgs:traingcn} is the process of GNN training (Algorithm \ref{alg:gcntrainingprocess}). Step \ref{state:datagenerationfortest} generates data by the agent and invokes the $\mathtt{Verifier}$ to test the topology. Step \ref{alga2cgs:testinggcn} is the procedure of testing GNN (Algorithm \ref{alg:testinggcnprocess}).
Since data from Step \ref{state:datagenerationfortest} are not used, they are used to examine the performance of a GNN classifier. 

\begin{algorithm}[htp]
\begin{algorithmic}[1]
\caption{$\mathtt{GNNTraining}$ algorithm}
\label{alg:gcntrainingprocess}
\STATE \textbf{Input:} GNN layer parameter as function $\tilde{f}_{\phi(t)}$, replay buffer $\mathcal{B}$, threshold value $q$.
\FOR{Epoch  $\tau =1,...,T$}
\STATE {Shuffle replay buffer $\mathcal{B}$ and collect $n$ pairs of data $\mathcal{D}_T=\{(s_i,r_i)\}_{i=1}^{n}$. Label data as $\{(s_i,\tilde{r}_i)\}_{i=1}^{n}$. $\forall (s_i,r_i)\in \mathcal{D}_T$, if $r_i\ge q, \tilde{r}_i=1$, else $\tilde{r}_i=0$.}
\STATE {For every $d_i=(s_i,r_i)\in \mathcal{D}_T$, update the function parameter $\phi(t)$ by optimizer.
}
\ENDFOR
\STATE \textbf{Return} New GNN layer parameter $\phi(t+1)\leftarrow\phi(t)$
\end{algorithmic}
\end{algorithm}
\vskip -0.2in

\begin{algorithm}[htp]
 \begin{algorithmic}[1]
 \caption{$\mathtt{GNNTesting}$ algorithm}
 \label{alg:testinggcnprocess}
 \STATE \textbf{Input}: {Data $\mathcal{D}=\{s^{j},a^{j},r^{j}\}_{j=1}^{n}$}, GNN $\tilde{f}_{\phi(t)}$, threshold value $q$.
 \STATE {Label data as $\{(s_j,\tilde{r}_j)\}_{j=1}^{n}$.} $\forall (s_j,r_j)\in \mathcal{D}$, if $r_j\ge q, \tilde{r}_j=1$, else $\tilde{r}_j=0$.
{
 \STATE Calculate $P(\tilde{f}_{\phi(t)}) = \sum_{j=1}^{n}\mathbb{I}(\tilde{f}_{\phi(t)}(s_j)=\tilde{r}_j)/n$. If $P(\tilde{f}_{\phi(t)})< 95\%$, $C_f=1$, else $C_f=2$. 
 \STATE \textbf{Return} $C_f$
}
 \end{algorithmic}
 \end{algorithm}
\vskip -0.1in

{The DRL-GS method possesses two features that can likely help its generalization capability. First, DRL-GS explicitly takes into consideration the network structure, which enables efficient searching for topology. 
Second, the topology verifier enables generating sufficient amount of training data, which supports the training and searching for DRL-GS. 
In the case study section below, we will demonstrate the superior performance of DRL-GS. }

\section{Case Study and Experiments}\label{chap:casestudy}
{In section \ref{chap:probdef}, we introduce the problem definition in an abstract form with generic
objective value function and abstract management requirements. In this section, we present a case study based on a real-world network topology optimization problem to demonstrate what  objective functions and constraints can be handled by our framework. We also discuss the difficulty of examining the validation of a topology under complex constraints and show how the DRL-GS algorithm outperforms heuristic methods.}

\subsection{The concrete topology optimization problem}\label{chap:concreteproblem}
In this section, we present a concrete case study based on a real-world network management scenario from China Mobile \cite{chinamobiledataset} using the information of a specific city. {Below, we will present the background description, problem definition, constraints and concrete DRL-GS implementation.} Table \ref{table:notationsofnodes} summarizes the key notations. In the case study, we use two datasets, the  small dataset and the large dataset.
In the small dataset, we use simulated location information to demonstrate the efficiency of compressing action space.
In the large dataset, all node locations are obtained from the real dataset. 
In both datasets, all network-related information, including node type, maximum utilization and average traffic volume,  are from the China Mobile dataset.

\vskip -0.2in
\begin{table}[ht]
\caption{Basic definitions of node features in the topology}
\label{table:notationsofnodes}
\centering
\vskip -0.1in
\renewcommand{\arraystretch}{1.2} 
\resizebox{0.5\textwidth}{!}{
\begin{tabular}{|l|l|ll}
\cline{1-2}
\textbf{Symbol}  & \textbf{Description} &   \\ \cline{1-2}
 $Type(v)$ & There are three node types, $T$, $H$ and $J$, representing\\
 & nodes in core layer, aggregation layer and access layer,\\ 
 & respectively. \\ \cline{1-2}
 $U_{\max}(v)$  & Maximum utilization of node $v$ with Gbps unit, repres-\\&enting maximum processing bandwidth of a node. %
 &  \\ \cline{1-2}
 $Pos(v)$  & Position of node $v$, i.e., the longitude and magnitude. &  \\ \cline{1-2}
 $Flow(v,t)$  & Average traffic volume passing through node $v$ per hour\\&with Mbps unit. &  \\ \cline{1-2}
\end{tabular}
}
\end{table}
\vskip -0.1in

We consider a link load balancing problem in the optical communication process at the bottom of the transmission network {among network elements, a physical infrastructure to serve thousands of users, i.e., nodes.} 
The objective is to optimize the topology structure from an initial topology, to achieve balance. Now we specify the  details of the $\mathtt{NetTopoOpt}$ formulation for the case study, including the objective function and constraints.

\subsubsection{Objective function}\label{chap:objectivefunction}
Recall that the objective function is $f(\bm{x})=U(\bm{x})+\gamma Cost(\bm{x},{\bm{x_0}})$, where $\bm{x}$ is the adjacent matrix of $G$, and $\bm{x_0}$ is its initial value. $Cost(\bm{x},{\bm{x_0}})$ is defined as:
\begin{equation}
\label{eq:computingcostdetails}
\begin{aligned}
Cost(\bm{x},{\bm{x_0}})&=\sum_{i,j}(\lambda_0 Dist(v_i,v_j)\mathbb{I}((\bm{x_0})_{ij}=0)\\
&+\lambda_1\mathbb{I}((\bm{x_0})_{ij}=1))| (\bm{x})_{ij}-(\bm{x_0})_{ij}|,
\end{aligned}
\end{equation} 
\noindent where $Dist(v_i,v_j)$ is the distance between two nodes $v_i$ and $v_j$;
$\mathbb{I}(\mathcal{E})$ is the indicator function of event $\mathcal{E}$, i.e., it is one if $\mathcal{E}$ is true. 
{$\lambda_0$ and $\lambda_1$ are fixed constant parameters. } This means that if an initial edge is not adapted, it does not influence the cost. If an edge is newly built, the cost is related to the distance. If an edge is removed, the cost is constant.

Next, we specify $U(\bm{x})$. Specifically, in the topology $G$, there exists a \emph{path set} denoted by $\mathcal{P}=\{1, 2, ..., P\}$, which consists of a set of $P$ paths described in Table \ref{tab:defofcalculationunits}. 
For each $p\in\mathcal{P}$, $Flow(p,t)=\sum_{v\in p\backslash\{Head(p),Tail(p)\}}$$Flow(v,t)$ denotes its bandwidth utilization at time $t$ (index hour) as the summation of average traffic volume for nodes in path $p$ except the end nodes. Then, 
the utility value $U(\bm{x})$ of a topology is defined as:
{
    \begin{align}
    \label{eq:computingdetails}
        U(\bm{x})&=\frac{1}{PT}\sum_{p\in \mathcal{P}}\sum_{t=1}^{T}\mathbb{I}\left(\left|\frac{Flow(p,t)}{B(t)}-1\right|\le\epsilon\right)\\
        &-\frac{1}{P}\sum_{p\in \mathcal{P}}(\alpha Sratio(p)+\beta Hratio(p)) \nonumber\\
        &-\frac{1}{T}\sum_{t=1}^{T}Var(\mathcal{P},t)-\max_{t}Var(\mathcal{P},t)-\min_{t}Var(\mathcal{P},t), \nonumber
    \end{align}
$Var(\mathcal{P},t)=\frac{1}{P}\sum_{p\in \mathcal{P}}\left(Flow(p,t)-\frac{1}{P}\sum_{p\in \mathcal{P}}Flow(p,t) \right)^2$ in Eq. \eqref{eq:computingdetails}, and $T=24$ hours, meaning that we calculate the utility value over a day.
} 
$Sratio(p)$ is the percentage of nodes in a sub path for a path, $Hratio(p)$ is the percentage of nodes with degree $1$ in topology $G$ for a path. 
$B(t)$ is the benchmark bandwidth utilization in time $t$, which is the average bandwidth utilization of paths in the initial topology,  and $\epsilon$ is the constant threshold value. 

\vskip -0.2in
\begin{table}[htp]
\caption{Definition of the notations for formation requirements}
    \label{tab:defofcalculationunits}
    \centering
    \vskip -0.1in
    \renewcommand{\arraystretch}{1.2} 
    \begin{tabular}{|l|l|ll}\cline{1-2}
       \textbf{Notations}  &\textbf{Definitions} \\\cline{1-2}
        Primary main path &The two end nodes of the path are both type-$T$\\&nodes and the intermediate nodes are type-$H$.\\\cline{1-2}
       Secondary main path &The two end nodes of the path are either $T$ or\\&$H$, and the intermediate nodes are of type-$J$. \\\cline{1-2}
       Main path  &Primary main path and secondary main path. \\\cline{1-2}
       Sub path &When there are multiple candidate secondary\\&main paths share the same two end nodes, one \\
       &of them will be declared the secondary main\\&path (see Algorithm \ref{alg:selectbestlink}), and the others will be \\
       &called sub paths.\\\cline{1-2}
       Path  &Main path and sub path. \\\cline{1-2}
       Hang node &Nodes with degree $1$ and connect to an inter-\\&mediate node in a path. \\\cline{1-2}
    \end{tabular}
\end{table}

{The first term $\frac{1}{PT}\sum_{p\in \mathcal{P}}\sum_{t=1}^{T}\mathbb{I}\left(\left|\frac{Flow(p,t)}{B(t)}-1\right|\le\epsilon\right)$ is the average of the indicator value for all links and times. It means that if $Flow(p,t)$ for a path is close to the benchmark $B(t)$, it contributes positive values. Multiple positive values mean that the majority of paths have similar bandwidth utilization to the benchmark, implying a load-balancing requirement on the path load. }
The second term $\frac{1}{P}\sum_{p\in \mathcal{P}}(Sratio(p)+Hratio(p))$ means that the value of sub ratio and hang ratio of a path cannot be too large. 
The last three terms represent the difference in bandwidth utilization among paths.  In the case study, the values are set to be  $\epsilon=0.4,\alpha=0.02,\beta=0.05,\lambda_0=10^{-8},\lambda_1=10^{-10},\gamma=-10^{-3}$. {One can see from the definition that the  objective function is nonlinear. Indeed, the function and constraints are both complicated functions of the  input variable $\bm{x}$, which significantly increases the difficulty of solving the problem.}

\subsubsection{Topology formation requirements}
Here we present the concrete constraints on network topology, i.e.,  \eqref{eq:con-phy} to \eqref{eq:con-feasibility} in $\mathtt{NetTopoOpt}$, for the case study.

{
\textit{The distance requirement}: The network requires that $\forall e=(v_i, v_j)$, $Dist(e)\le D_{max}$, where $Dist(e)$ is the length of the link, and $D_{max}$ is the maximum allowed distance. 
}

\textit{The utility requirement}: { For any path $p\in \mathcal{P}$, $Flow(p,t)\le L_{max}=1$, where $L_{max}$ is the maximum capacity value of a path.} Notation $Flow(p,t)$ replaces $Load(e)$ described in Section \ref{chap:probdef} and we examine $Flow(p,t)$ to state whether or not a topology satisfies utility requirement.

\textit{The feasibility requirement}: The topology $G$ needs to be connected, which guarantees  the connectivity of any two nodes.  
{For any path $p\in\mathcal{P}$, the number of nodes in path $p$ cannot exceed $15$ as physical requirements. Note the number $15$ comes from the real-world problem, and our framework can easily handle other path-length limit values. Any type-$H$ node must belong to one primary main path to prevent duplication in multiple primary main paths. }
For a path $p$, we denote  the two end nodes, by  $v_h$ and  $v_t$. For other nodes $v\in p$, the maximum utilization value of node $v$, denoted as $U_{\max}(v)$, cannot be larger than the maximum value of the end nodes, i.e.,  $U_{\max}(v)\le  \max\{U_{\max}(v_h),U_{\max}(v_t)\}$. {If $p$ is a secondary main path, the two end nodes should have the same maximum utilization value, i.e., $U_{\max}(v_h)=U_{\max}(v_t)$.}
There can also be nodes that connect to intermediate nodes in a path and has a degree $1$. We call such nodes hang nodes. Then, for a hang node $v_H$ attached to another node $v_A$, $U_{\max}(v_H)\le U_{\max}(v_A)$. These requirements  prevent overloading in a path.

\subsection{DRL-GS for topology search}
For optimizing the topology, we use the DRL-GS algorithm as a search method. Recall that the DRL-GS algorithm consists of two main parts: the verification part to validate whether or not a topology can calculate the objective value, and the optimization part by using RL algorithm to search topologies for optimizing the function. We specify the steps of DRL-GS in this case.

\subsubsection{The $\mathtt{Verifier}$ Algorithm}
Based on constraints  \eqref{eq:con-phy} to \eqref{eq:con-feasibility}, 
we construct the topology validation method $\mathtt{Verifier}$ shown in  Algorithm \ref{alg:verification}, which determines if a generated topology is validated and calculates its objective value. 
If a topology is invalid, the algorithm returns false and a low objective value (set to $-10$ in our implementation). Otherwise, it returns true and the objective value. 
In Algorithm \ref{alg:verification}, $N(p)$ is the number of nodes in path $p$. $Head(p)$ and $Tail(p)$ are the end nodes of a path. $Dist(e)$ is the distance of an edge $e$. {$Deg(v,G)$  is the degree of a node $v$ in topology $G$. $Nbr(v,G)$ denotes the set of neighbor nodes of node $v$ in $G$.}

\begin{algorithm*}[htp]
\begin{algorithmic}[1]
\caption{The $\mathtt{Verifier}$ algorithm}
\label{alg:verification}

\STATE {\textbf{Input:} Topology $G=(V,E)\ (E=E(\bm{x})),D_{max}$ and features of a node: $U_{\max}(V)$, $Type(V)$.}
\STATE {Generate a sub-topology $G' = (V',E')\subset G$. $V'=\{v|v\in V,Type(v)=T/H\}$, $E'=\{e|e=(v_i,v_j)\in E,v_i,v_j\in V'\}$.  $\bm{I}_{31}=\neg(\cup_{v\in V'}\mathbb{I}(Type(v)=H \land Deg(v,G')\neq 2))$. \label{ver:bfsmain} }
\STATE {Run $\mathcal{P}_P=\mathtt{BFS}(G')$ to search all primary main paths $\mathcal{P}_P = \{p_P^{(i)}\}_{i=1}^{m}$ between nodes $v\in V',Type(v)=T$.
\label{ver:bfsmain2} }
\STATE {Generate a sub-topology $G''=(V'',E'')\subset G$, $V''=V$, $E''=\{e|\forall p_P^{(i)}\in\mathcal{P}_P,e\notin p_P^{(i)}\}$. \label{ver:bfssub} }
\STATE { Run $\mathcal{P}_S=\mathtt{BFS}(G'')$ to search all possible secondary main paths $\mathcal{P}_S = \{p_S^{(i)}\}_{i=1}^{n}$ between nodes $v\in V'',Type(v)=T/H$.\label{ver:bfssub2} }
\STATE {Run $\mathcal{P}_S^{0},\mathcal{P}_S^{1},\mathcal{RL},\bm{I}_{32}=\mathtt{SelectBestPath}(\mathcal{P}_S)$ (Algorithm \ref{alg:selectbestlink}) to output $\mathcal{P}_S^{0} = \{p_S^{(i)}\}_{i=1}^{n_0}\subset \mathcal{P}_S$ as secondary main path set, $\mathcal{P}_S^{1} = \{p_S^{(j)}\}_{j=1}^{n_1}\subset \mathcal{P}_S$ as the sub path set, attached relationship $\mathcal{RL}=\{(p_S^{(i)},p_S^{(j)})\},p_S^{(i)}\in \mathcal{P}_S^{0},p_S^{(j)}\in \mathcal{P}_S^{1}$ and $\bm{I}_{32}$. \label{vef:choosebestlink}}
\STATE {$\bm{I}_1=\cap_{e\in E}\mathbb{I}(Dist(e)\le D_{max})$. (Constraints (\ref{eq:con-phy}))} \label{vef:distance}\\
{ $\bm{I}_{2}=\cap_{ t,p\in \mathcal{P}_S\cup \mathcal{P}_P} \mathbb{I}(\sum_{v\in p}Flow(v,t)\le\max\{U_{\max}(Head(p)),U_{\max}(Tail(p))\}).$\label{vef:subutility} (Constraints (\ref{eq:con-load}))}
\STATE { $\bm{I}_{33}=\cap_{ p\in \mathcal{P}_S\cup \mathcal{P}_P,v\in p} \mathbb{I}(U_{\max}(v)\le\max\{U_{\max}(Head(p)),U_{\max}(Tail(p))\})$,\\
$\bm{I}_{34}=\cap_{p\in \mathcal{P}_S^{0}\cup \mathcal{P}_P}\mathbb{I}(N(p)\le 15)$, $\bm{I}_{35}=\cap_{v\in V}\mathbb{I}(Deg(v,G)\neq 0)$,\\
$\bm{I}_{36}=\cap_{v_1\in V,Deg(v_1,G)=1,v_2\in Nbr(v_1,G)}\mathbb{I}(U_{\max}(v_1)\le U_{\max}(v_2)),\bm{I}_3=\cap_{j}\bm{I}_{3j}$. \label{vef:sublength} (Constraints (\ref{eq:con-feasibility}))}
{
\IF {$\cap_{j}\bm{I_j}=True$}
\STATE \textbf{Return} $\cap_{j}\bm{I_j}$,$f(\bm{x})=\mathtt{CalObjValue}(\mathcal{P}_P,\mathcal{P}_S^{0},\mathcal{P}_S^{1},\mathcal{RL},G)$ 
\ELSE 
\STATE \textbf{Return} $\cap_{j}\bm{I_j}$,$f(\bm{x})=-10$
\ENDIF
}
\end{algorithmic}
\end{algorithm*}

{Step \ref{vef:distance} is the distance constraints and the utility constraints. The other steps are the feasibility requirements as follows.
Step \ref{ver:bfsmain} ensures that the degree of type-$H$ nodes is $2$ in the sub-topology. Step \ref{vef:choosebestlink} examines if the secondary main path candidates satisfy the requirements by executing the $\mathtt{SelectBestPath}$ algorithm. Step \ref{vef:sublength} checks these feasibility requirements: the number of nodes in the main path is within $15$; the maximum utilization value of the intermediate nodes is no  larger than the end nodes; topology $G$ is connected; the maximum utilization of the hang node is no larger than that of the connected node.
}

In order to verify the feasibility and compute the objective value, in Steps \ref{ver:bfsmain2} and \ref{ver:bfssub2}, we run breadth-first search algorithm ($\mathtt{BFS}$) to search all possible connected paths between two nodes. However, we cannot determine which is the secondary main path if there are many candidates. 
The $\mathtt{SelectBestLink}$ algorithm (Algorithm \ref{alg:selectbestlink}) helps the determination. 
The overall verification steps help determine whether or not a topology is valid. {For an invalid topology, the algorithm returns false with a large negative value $-10$. Otherwise, the $\mathtt{Verifier}$ executes the $\mathtt{CalObjValue}$ algorithm to compute the objective value as described in Algorithm \ref{alg:calobjvalue}. }

\begin{algorithm*}[htp]
\begin{algorithmic}[1]
\caption{$\mathtt{SelectBestPath}$ algorithm}
\label{alg:selectbestlink}
\STATE \textbf{Input}: All possible secondary main paths $\mathcal{P}_S = \{p_S^{(i)}\}_{i=1}^{n}$
\STATE { Initialize a secondary main path $\mathcal{P}_S^{0}$, a sub path $\mathcal{P}_S^{1}$ and an attached relationship buffer $\mathcal{RL}$.}
\STATE { Generate a head-tail node buffer $\mathcal{BN}=\{\{Head(p_S^{(i)}),Tail(p_S^{(i)})\}|p_S^{(i)}\in \mathcal{P}_S\}$ and a path buffer $\mathcal{BP}=\{\mathcal{BP}(v_i,v_j)\}$ ($\mathcal{BP}(v_i,v_j)=\{p|p\in \mathcal{P}_S,\{Head(p),Tail(p)\}=\{v_i,v_j\}\}$).}
\STATE { $\forall \{v_i,v_j\}\in\mathcal{BN}$, take $\mathcal{BP}^{f}(v_i,v_j)=\{p|p\in \mathcal{BP}(v_i,v_j),\exists v_k,v_l\in p, Type(v_k)=Type(v_l)=J,U_{\max}(v_k)\neq U_{\max}(v_l) \},\mathcal{BP}^{s}(v_i,v_j)=\mathcal{BP}(v_i,v_j)\backslash\mathcal{BP}^{f}(v_i,v_j),\mathcal{P}_S^{1}\leftarrow\mathcal{P}_S^{1}\cup\mathcal{BP}^{f}(v_i,v_j)$.}
\STATE {$\bm{I}_{32}=\cap_{\{v_i,v_j\}\in\mathcal{BN}}\mathbb{I}(\mathcal{BP}^{s}(v_i,v_j)\neq\emptyset)$. (Constraint (\ref{eq:con-feasibility})).}
\STATE {$\forall \{v_i,v_j\}\in\mathcal{BN}$, take $p_{\{v_i,v_j\}}\in\mathcal{BP}^{s}(v_i,v_j) $ that $\forall p\in \mathcal{BP}^{s}(v_i,v_j),N(p)\le N(p_{\{v_i,v_j\}})$.}
\STATE {$\forall \{v_i,v_j\}\in\mathcal{BN}$, $\mathcal{P}_S^{0}\leftarrow \mathcal{P}_S^{0}\cup \{p_{\{v_i,v_j\}}\}, \mathcal{P}_S^{1}\leftarrow\mathcal{P}_S^{1}\cup(\mathcal{BP}^{s}(v_i,v_j)\backslash\{p_{\{v_i,v_j\}}\}),\mathcal{RL}\leftarrow\mathcal{RL}\cup\{(p_{\{v_i,v_j\}},p')\},\forall p'\in \mathcal{BP}^{s}(v_i,v_j)$. (Find secondary main paths in candidates).}
\STATE {\textbf{Return} $\mathcal{P}_S^{0}$, $\mathcal{P}_S^{1}$ ,$\mathcal{RL}$, $\bm{I}_{32}$.}
\end{algorithmic}
\end{algorithm*}

\begin{algorithm*}[htp]
\begin{algorithmic}[1]
\caption{$\mathtt{CalObjValue}$ algorithm}
\label{alg:calobjvalue}
\STATE {\textbf{Input}: The primary main path set $\mathcal{P}_P=\{p_P^{(i)}\}_{i=1}^{m}$, secondary main path $\mathcal{P}_S^{0} = \{p_S^{(i)}\}_{i=1}^{n_0}$, sub path set $\mathcal{P}_S^{1} = \{p_S^{(j)}\}_{j=1}^{n_1}$, the attached relationship $\mathcal{RL}=\{(p_S^{(i)},p_S^{(j)})|p_S^{(i)}\in \mathcal{P}_S^{0},p_S^{(j)}\in \mathcal{P}_S^{1}\}$. The topology $G$ that generates the paths. }
\STATE  {$\forall p_S^{(i)}\in \mathcal{P}_S^{0}$, initialize $Len(p_S^{(i)})=0$.}
\STATE  {$\forall (p_S^{(i)},p_S^{(j)})\in\mathcal{RL}$, take $Len(p_S^{(i)}) \leftarrow Len(p_S^{(i)}) +  N(p_S^{(i)}\cup p_S^{(j)})-N(p_S^{(i)}),p_S^{(i)}\leftarrow p_S^{(i)}\cup p_S^{(j)}$.}
\STATE {$\forall p\in \mathcal{P}_S^{0}\cup \mathcal{P}_P$, $  Flow(p,t)=\frac{\sum_{v\in p\backslash\{Head(p),Tail(p)\}}Flow(v,t)}{\max\{U_{\max}(Head(p)),U_{\max}(Tail(p))\}}$
,  $Hratio(p)=\frac{\sum_{v\in p}\mathbb{I}(Deg(v,G)=1)}{N(p)}$. $\forall p_S^{(i)}\in \mathcal{P}_S^{0}$, $Sratio(p_S^{(i)}) = \frac{Len(p_S^{(i)})}{N(p_S^{(i)})}$. (Fundamental units to calculate $f(\bm{x})$).}
\STATE {\textbf{Return} $f(\bm{x})=U(\bm{x})+\gamma Cost(\bm{x},\bm{x_0})$. (Equation (\ref{eq:computingcostdetails}),(\ref{eq:computingdetails}))}
\end{algorithmic}
\end{algorithm*}
We now use Fig. \ref{fig:example_network_planning} as an example to instantiate the above steps. In this example, the distance of the edge between node $v_1$ and $v_7$ cannot be larger than $500$m. Consider a path $p=\{v_1\mbox{-}v_7\mbox{-}v_8\mbox{-}v_2\}$, $Flow(p,t)$ is bounded by the maximization utility $1$. 
The feasibility requirements are as follows. The topology should be connected. Any path cannot contain more than $15$ nodes. 
For instance, the path $p$ above with $4$ nodes satisfies the requirement.
For the path $p$ above, the maximum utilization value of nodes $v_7$ and $v_8$ cannot be larger than those of nodes $v_1$ and $v_2$ (the end nodes). The maximum utilization value of node $v_{21}$ cannot be larger than that of node $v_{20}$. 

\subsubsection{Action Compression in DRL-GS}\label{chap:actioncompression}
Although DRL enables efficient search, it still suffers from the curse of dimensionality when facing problems with high-dimensional action spaces. Consider a network topology in our case with $23$ nodes and $72$ valid edges. This leads to an action space with $2^{72}=4.7\times10^{21}$ options if we use a brute force enumeration, which is almost impossible to search over. To solve this problem, we propose a compact definition of the action space in our RL algorithm. 
Specifically, we define a compact action consisting of five steps, 
each corresponding to a decision at a particular level (see below), which will significantly improve the search efficiency of the algorithm. 

\begin{figure}[ht]
    \begin{center}
    \includegraphics[width=0.5\textwidth]{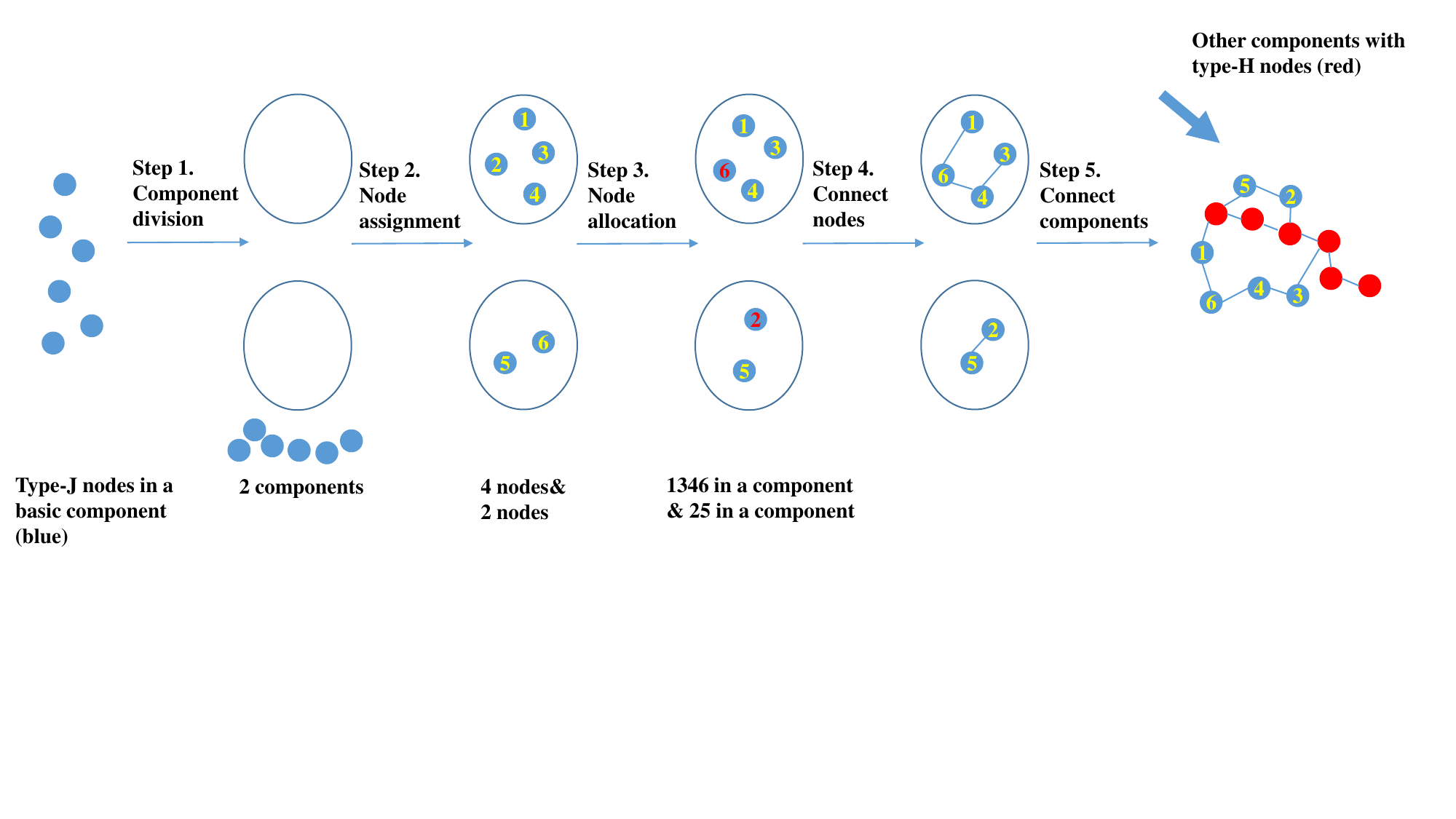}
    \vskip -0.7in
    \caption{Schematic diagram of action compression in five steps.}
    \end{center}
    \label{fig:actioncompression}
\end{figure}

Before introducing the steps, we give some basic definitions first.
We generate a connected topology $G^*=(V,E^*)$ that we connect all  edges satisfying $Dist(e)\le D_{max}$. 
Based on the topology $G^*$, we generate sub-topologies $G_1$ and $G_2$. $G_1$ consists of all  type-$H$ nodes and all  links connecting type-$H$ nodes in topology $G^*$. 
$G_2$ is alike $G_1$ with type-$J$ nodes format.
We run the $\mathtt{BFS}$ algorithm for $G_1$ and $G_2$ to generate all connection components
regarded as basic components preparing for generating new components. Based on these components, 
we define our five-step action as below.
Fig. \ref{fig:actioncompression} shows how to utilize these steps to generate a topology.
\vskip -0.1in

\textit{Step 1.} The first step decides how many  sub-components to have from a basic  component, e.g., we can divide a basic component with $6$ nodes into $2$ sub-components in Fig. \ref{fig:actioncompression}. 

\textit{Step 2.} The second step decides how many nodes to assign to each sub-component. 
For example, consider the $2$ sub-components obtained from the first step in Fig. \ref{fig:actioncompression}.  There are $3$ potential choices to assign nodes, i.e., $\{(5,1),(4,2),(3,3)\}$. 
Specifically, $(4,2)$ means that one sub-component has $4$ nodes and the other has $2$ nodes.  

\textit{Step 3.} The third step decides the allocation of nodes for each sub-component. Consider the example $(4,2)$ for node assignment, there are $15$ choices to allocate $6$ different nodes into two sub-components. Note that the first three steps influence the objective value, by affecting  the bandwidth utilization of a path. {For a component with $m$ nodes, if divided into $k$ sub-components, the number of overall choices in step $1\sim 3$ is a second Stirling number $S(m,k)$.}

\textit{Step 4.} The fourth step connects the nodes in the sub-components to make each sub-component fully connected. { For a sub-component with $m$ nodes, the number of choices is at most $O(2^{\frac{m(m-1)}{2}})$. } For instance, in Fig. \ref{fig:actioncompression}, the sub-component with $4$ nodes have at most $2^{\frac{4\times 3}{2}}=64$ choices. This step influences the objective value by determining the sub ratio and hang ratio of a path.

\textit{Step 5.} In the fifth step, we connect all the sub-components according to the management requirements to form one connected component.   
{In the case of connecting two sub-components where one has $m$ type-$H$ nodes and the other has $m'$ type-$J$ nodes, the number of choices is at most $m(m-1)$.}
The step influences the objective value by determining the maximum utilization of the end nodes in a path to influence the bandwidth utilization.

By using the above five-step action, we obtain a  compact action space compared to working on modifying each link. 
However, the resulting action space is still very large. For instance, a topology has only one component with $16$ nodes initially. In the first step, we divide the component into $4$ sub-components. In the second step, we decide that every sub-component has $4$ nodes. {Based on the first and second steps, the third step has $2.6$ million potential choices (the number of possible ways to allocate $16$ different nodes evenly  into $4$ components  is $16!/(4!)^5$) while the fourth step has $3.1$ million potential choices (for each component with $4$ nodes, the number of all possible choices for connecting the component is $42$, and for all components the number is $42^{4}$).} It is impossible for the RL agent to train a policy in this situation.

To further tackle the curse of dimensionality, in actual implementation, we further reduce the action space by imposing restrictions on the choices in different steps, based on prior knowledge of the setting. 
{First, we can impose some constraints from the prior knowledge to restrict the step set without using $\mathtt{Verifier}$, e.g., we can restrict step $4$ in one action to satisfy the distance requirements. Second, we can utilize the data information as prior knowledge to restrict the steps, e.g., we may restrict step $2$ to a few choices based on the traffic volume, because these choices may have better objective values. Specifically, if there are too many choices in a step, we restrict ourselves to a pre-specified option set so that the elements in the set are guaranteed enough to be a valid topology and the objective value of these elements is probably good enough compared to the abandoned actions.} The elements in the step set are chosen according to prior knowledge such that the size of the step set is neither too large nor too small, to enable efficient decision making. 
Although this restriction may limit possible actions, it is effective in reducing the size of the action space and allows for higher learning efficiency. 

\subsubsection{Reward function representation}
\begin{figure}[htp]
    \centering
    \includegraphics[width=0.5\textwidth]{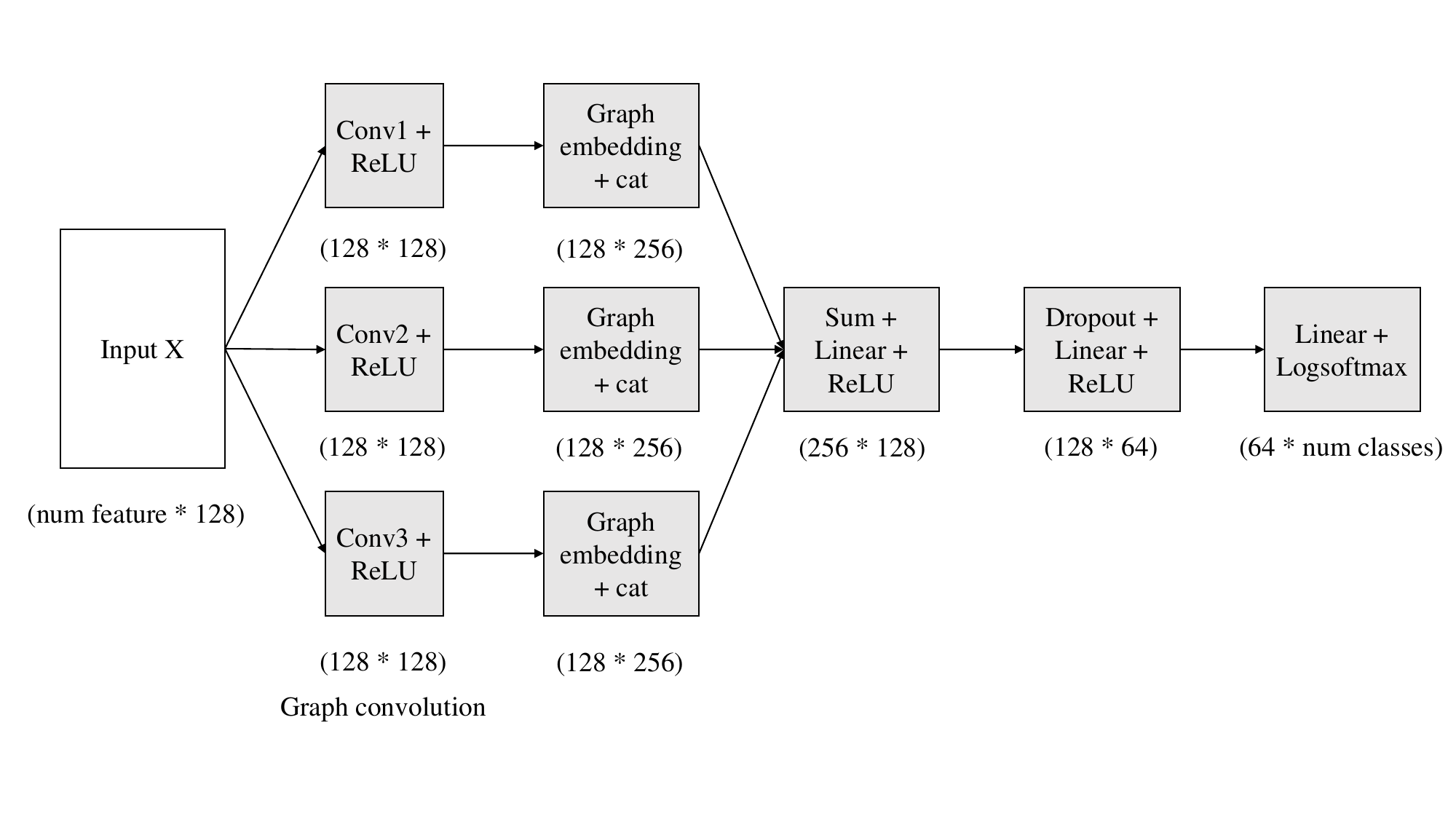}
    \vskip -0.3in
    \caption{Implementation details in GNN for classification.}
    \label{fig:gcnlayer}
\end{figure}

To enable efficient search, we introduce a graph neural network (GNN) to facilitate reward learning. The motivation is that in large networks, the $\mathtt{Verifier}$ algorithm is time-consuming,
while GNN can classify a good topology efficiently. The reason is that GNN contains nonlinear components to extract the feature and generate the discrimination directly, while the $\mathtt{Verifier}$ algorithm judges the validation and calculates the objective value by generating many paths.
The structure of the GNN is shown in Fig. \ref{fig:gcnlayer} based on Pytorch Geometric \cite{fey2019fast} for implementation. It consists of $3$ parallel graph convolution networks, global mean and maximum pooling as the graph embedding, $3$ linear layers with activation functions and a dropout layer and output a log softmax value for judgment.

\vskip -0.15in
\vskip -0.1in

Next, we specify the training of a GNN. We first generate a set of valid topologies manually by running $\mathtt{Verifier}$. {We then set a value threshold $q$ (a hyperparameter determined by grid search to balance GNN training and RL training)} 
and label the generated topology a good one (with value $1$) or a poor one (with value $0$) based on whether the objective value is above or below the threshold.
We then utilize these labeled topologies to train a classifier by GNN designed above. 
{The input of the GNN is the node information that the node type, maximum utilization, position and bandwidth utilization in time $t= 1, 2, ..., 24$, and the adjacent matrix of the topology. The output is a log softmax value as the probability for classification.
For the convolution layer of GNN}, the overall activation function is: $ H(l+1) = \sigma(D^{-1/2}(\bm{x}+I)D^{1/2}H(l)W(l))$.
Here $H(l)$ is the $l$-th convolution layer input and $H(0)=X$ is the input.  
$\bm{x}$ is the adjacency matrix of topology $G$, $I$ is the identity matrix, $D$ is the degree matrix of topology $G$. $W(l)$ is the parameter in the $l$-th layer. $\sigma(\cdot)$ is the activation function for non-linearity in GNN.

\subsubsection{DRL framework implementation}
{In the DRL framework, we use the Advantage Actor-Critic (A2C) and Proximal Policy Optimization (PPO) algorithms \cite{mnih2016asynchronous,schulman2017proximal} to learn the RL agent. Note that our framework is general and other suitable DRL algorithm can also be applied. 
The RL agent uses the policy $\pi_\theta(a|s)$ to search topologies as an actor shown as a probability distribution.
The critic $V^{\pi}(s)$ utilizes the reward calculated by the initial objective value from $\mathtt{Verifier}$ algorithm or the label value of GNN for simplification to evaluate the performance of a topology.
We utilize multi-layer perceptrons as the basic structure of parameterized policy network and value network under the stable baselines3 framework \cite{raffin2021stable} to help finish implementation and train the agent.
The key information and hyper-parameters are shown in Table \ref{tab:parameters}. }

\vskip -0.1in
\begin{table}[htp]
\caption{ {Key information and hyper-parameters in RL training. }}
    \label{tab:parameters}
    \centering
    \renewcommand{\arraystretch}{1.4} 
    \begin{tabular}{|l|l|l|}
\cline{1-3}
\textbf{Algorithms}  & \textbf{Policy/Value Network} & \textbf{Learning rate}  \\ \cline{1-3}
A2C & $64$, tanh, $64$, tanh, Linear & $7\cdot10^{-4}$ \\ \cline{1-3}
{PPO} & {$64$, tanh, $64$, tanh, Linear} & {$3\cdot10^{-4}$}  \\ \cline{1-3}
\textbf{Algorithms}  & \textbf{Timesteps}  &\textbf{Discount factor}  \\ \cline{1-3}
A2C&$2\cdot10^6$/$10^6$/$10^5$ &0.99 \\ \cline{1-3}
{PPO} &{$2\cdot10^6$/$10^6$} &{0.99} \\ \cline{1-3}
    \end{tabular}
\end{table}

In RL training, the algorithm first initializes the parameters of the policy. For each epoch, based on the policy, several actions are sampled corresponding to new topologies. By repeatedly choosing from the sampled actions for $H$ steps, the critic finally finds the best topology that gives the highest reward. At the end of the epoch, we compute the policy gradient loss (entropy loss) and critic gradient loss (value loss) and update the parameters. The entropy loss and value loss derived by the gradient loss are regarded as representatives of exploration and exploitation metrics.

{
\subsubsection{DRL-GS algorithm complexity}
The complexity of the algorithm steps are mainly due to executing the $\mathtt{Verifier}$, GNN classifier and RL training. In the case study, the computational complexity of the $\mathtt{Verifier}$ is $O(|V|^2)$ ($|V|$ is the number of vertices) because of finding all secondary main paths. However, using the GNN takes $O(|V|)$ computational complexity, which improves efficiency. Since DRL-GS is an iterative algorithm, the overall complexity of the algorithm also depends on the number of iterations required for convergence. Using the Adam optimizer for our training needs $O(1/\epsilon^2)$ the number of steps from optimization theory \cite{kingma2014adam}.}

\subsection{Benchmark Comparison and Experiments}

In the experiment, we test the performance of DRL method with two different datasets specified in Section \ref{chap:concreteproblem}. 
%
In order to show that RL policy helps search a topology with satisfied objective value, we compare DRL-GS with an one-step optimization algorithm shown in Algorithm \ref{alg:onestep}, which is often adopted in practice by utilizing prior knowledge of human experts for heuristic topology search, as a baseline method. Besides, 
we compare DRL-GS with random policy (randomly choosing an action in the full action space to generate a topology by randomly building or removing a few links) as another basic method to show that RL policy learns the ability to optimize the topology. 

\subsubsection{Small dataset}\label{exper:smalldata}

In the small dataset, the topology has $8$ nodes ($2T$+$3H$+$3J$) with information explained in Section \ref{chap:concreteproblem}. The initial topology has $11$ edges. The maximum distance of the connected edge is $D_{max}=500$m. Thus, $20$ edges can be connected in total.  
We use A2C-GS and PPO-GS to train the agent under the full space with $2^{20}$  choices and compressed space with $10$ combinations 
using the method in Section \ref{chap:actioncompression}. 
Initially, there are $2$ components that all nodes are type-$H$ (one has $2$ nodes and one has $1$ node) and $1$ component that all nodes are type-$J$ with $3$ nodes. The overall choices of step $1\sim3$ are that,
for the component with $2$ nodes,
there are $2$ choices and for the component with $3$ nodes, there are $5$ choices. In the fourth step and fifth step, we use  one choice that guarantees  the management requirements. 

\vskip -0.2in
\begin{figure}[htp]
    \centering
    \includegraphics[width=0.5\textwidth]{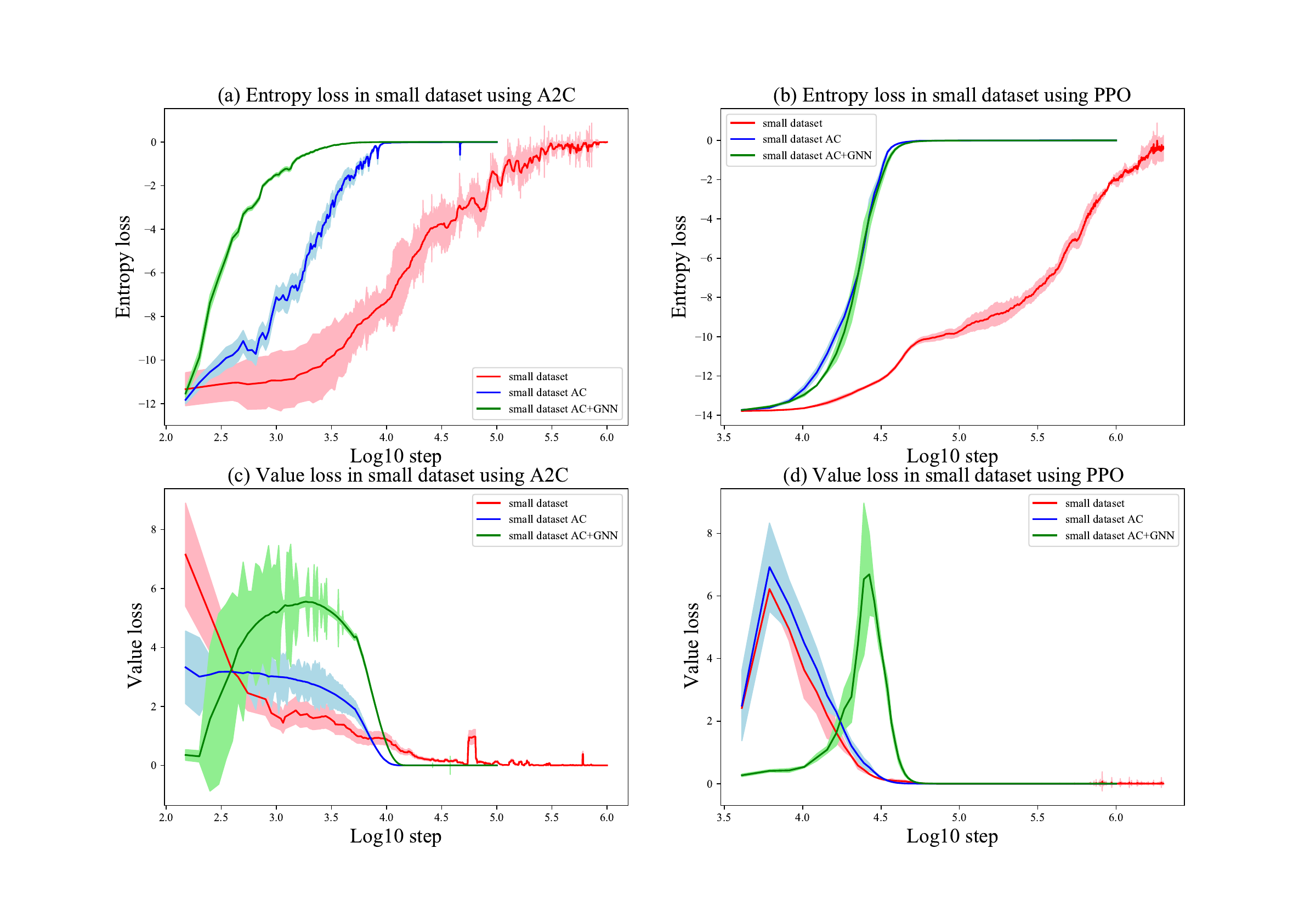}
    \vskip -0.2in
    \caption{{Entropy loss (a,b) and value loss (c,d) of RL training in small dataset using A2C (a,c) and PPO (b,d). The red loss curve captures the agents trained in the full space. The blue loss curve captures the agents trained in compressed space (AC means action compression). The green loss curve captures the agents trained in compressed space with GNN classifier.}
    }
    \label{fig:trainingRLsmalldataset}
\end{figure}
\vskip -0.1in

{Fig. \ref{fig:trainingRLsmalldataset} shows the entropy loss and value loss of training an RL agent in the small dataset, which demonstrates convergence in different training settings. 
Specifically, training the A2C agent in the original full space requires $10^6$ steps for convergence, while it only needs 
$5\cdot10^4$ steps to guarantee the convergence of training in our reduced sub-space. For the PPO agent, it requires $2\cdot 10^6$ steps for convergence in the full space, while $10^5$ steps are enough in compressed space.}

\begin{figure}[htp]
    \centering
    \includegraphics[width=0.5\textwidth]{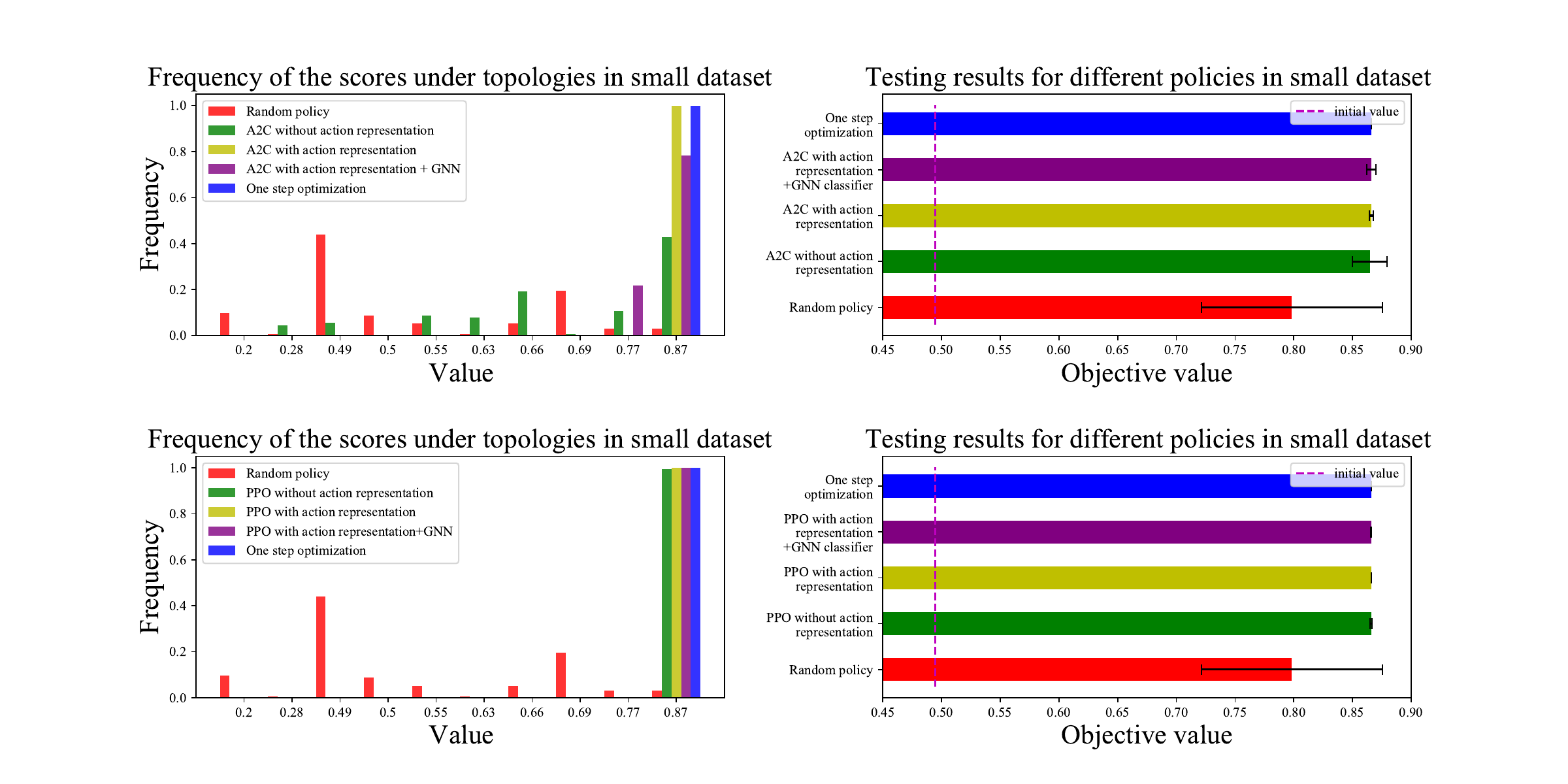}
    \vskip -0.2in
    \caption{{Left: Frequency of the scores under topologies found by different policies in small dataset. Right: Testing results for different policies in small dataset. } }
    \label{fig:frequency_smalldata}
\end{figure}

The left figure in Fig. \ref{fig:frequency_smalldata} shows the frequency of the scores by different methods. 
The frequencies are computed as follows.
First, initialize the environment with a random topology and search one topology by the policy, calculate the score of the topology, repeat the process for $5\cdot10^4$ times and compute the frequency of the scores. 
{
The fraction of time that the A2C agent trained in full space finds the best topology is roughly $40\%$ and for PPO agent, it is over $99\%$ (green). Both are much higher than that of the random policy (red, less than $5\%$). It shows that DRL-GS finds the best topology with a higher success rate than the random policy without any prior knowledge. Moreover, the fractions that the DRL agents trained in compressed space find the best topology are both nearly $100\%$ (yellow), which shows that by action compression, the DRL agent achieves a better performance. %
We note that the one-step optimization also finds the optimal topology (blue). This is  due to the small scale of the dataset. 
We will see later that in a larger scale experiment, DRL-GS outperforms one-step optimization significantly.
}

\vskip -0.2in
\vskip -0.1in

The right figure in Fig. \ref{fig:frequency_smalldata} shows the testing results of different methods, with  the mean values shown in colored bars, and the standard deviations shown in black error bar. 
Specifically, for 
a given random topology, we use the policy to search $5$ topologies and pick the maximum objective value among them. We repeat the process for $5\cdot10^4$ times, and calculate the mean value and standard deviation. 
{We see that both the A2C policy (green, $0.8648\pm 0.0148$) and PPO policy ($0.8658\pm 0.0008$) perform better than random policy (red, $0.7984\pm 0.0770$).  
In addition, A2C policy trained in compressed space further improves the performance (yellow, $0.8660\pm 0.0017$) while no better than one step optimization (blue, $0.8660\pm 0.0001$). However, PPO policy trained in compressed space performs as well as one step optimization (yellow, $0.8660\pm 0.0001$). }

\vskip -0.1in
\begin{figure}[htp]
    \centering
    \includegraphics[width=0.5\textwidth]{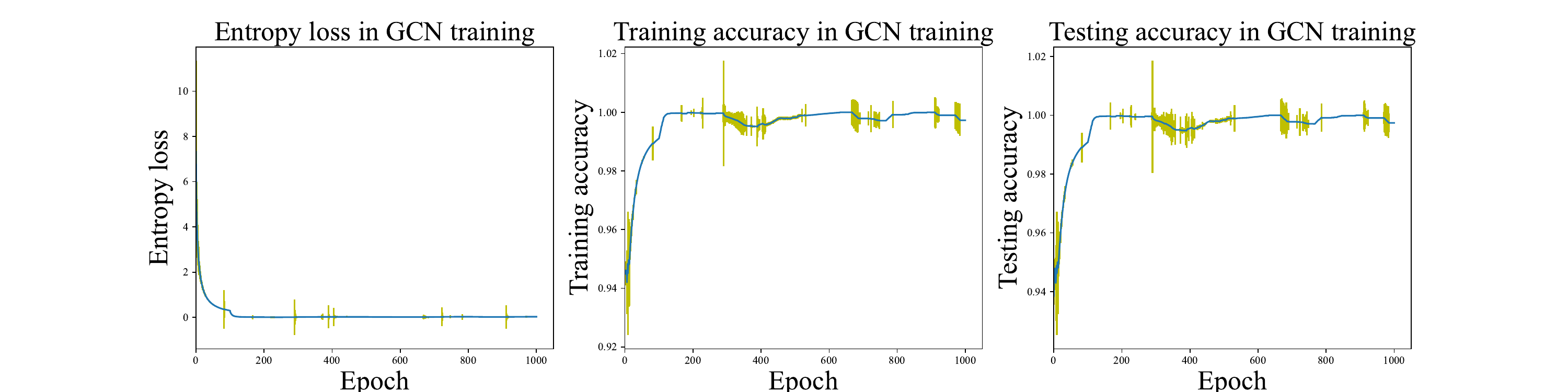}
    \vskip -0.1in
    \caption{GNN training with entropy loss, training accuracy and testing accuracy in small dataset.}
    \label{fig:gcntraining}
\end{figure}
\vskip -0.1in

Next we train RL agent in compressed space using GNN and examine the training and testing accuracy of the GNN. 
We use the Adam optimizer with the learning rate $0.005$, $q=0.8$ and training $10^3$ epochs.
Fig. \ref{fig:gcntraining} shows the results of the GNN training with the average value and the standard deviation of the entropy loss, training accuracy and testing accuracy.
The entropy loss converges to $10^{-4}$, the training accuracy and testing accuracy is higher than $99\%$, which shows that the GNN can classify a topology in high accuracy.

{
The right figure in Fig. \ref{fig:frequency_smalldata} also shows the testing results of the RL agent trained in compressed space adding GNN classifier with similar mean value and higher standard deviation (purple, A2C: $0.8660\pm 0.0039$, PPO: $0.8660\pm 0.0003$).
}
In the small dataset case, since  the topology structure can be verified  easily by the $\mathtt{Verifier}$ algorithm, GNN does not show its full advantage in RL training, and algorithms need roughly $10$ hours to train policy with action compression. However, we will see in the large-scale dataset that GNN can greatly reduce the running time.

\subsubsection{Large dataset}

In the large dataset, the topology has $23$ nodes ($2T$+$5H$+$16J$) $33$ edges with information explained in Section \ref{chap:concreteproblem}.
The maximum distance of the connected edge is $5$km with a total of $72$ edges connectable. 
Therefore, the number of the full action space is $2^{72}=4.7\times 10^{21}$, in which case the agent suffers from the curse of dimensionality, and it is critical for the agent to learn in our compressed action sub-space for higher efficiency.

The action sub-space is defined 
in Section \ref{chap:actioncompression}. 
In order to show the trade-off between efficiency and performance, we conduct experiments in two different action sub-spaces, one is smaller with fewer possible actions while another one is larger.
We illustrate our design of the two action sub-spaces as below.
Initially, there are two components, one with $5$ type-$H$ nodes and another with $16$ type-$J$ nodes. In the large space case,
we do not restrict the possible actions for the first basic component. For the second basic component with $16$ nodes, consider the first action step with the number of sub-components $m_1$,
if we choose $m_1=1$, we restrict that there is one choice in step $2\sim5$ that guarantees the management requirements.
Instead, if we choose $m_1=2$, we restrict the second step in the set $\{(15,1),(14,2)\}$ (example: $(15,1)$ means a choice in the second step that one component has $15$ nodes and another has $1$ node).
We have no compression on step $3$ and we restrict that there is one choice in step $4\sim5$ that guarantees the management requirements too. 
The design of small space has one difference that if we choose  $m_i=2$ in the first step for the component with $16$ nodes, we restrict that there is one choice $(15,1)$ in the second step. 
The design helps implementation of training an RL agent and allows us to compare the performance of RL agent trained under different sizes of action spaces.

\begin{figure}[H]
    \centering
    \includegraphics[width=0.5\textwidth]{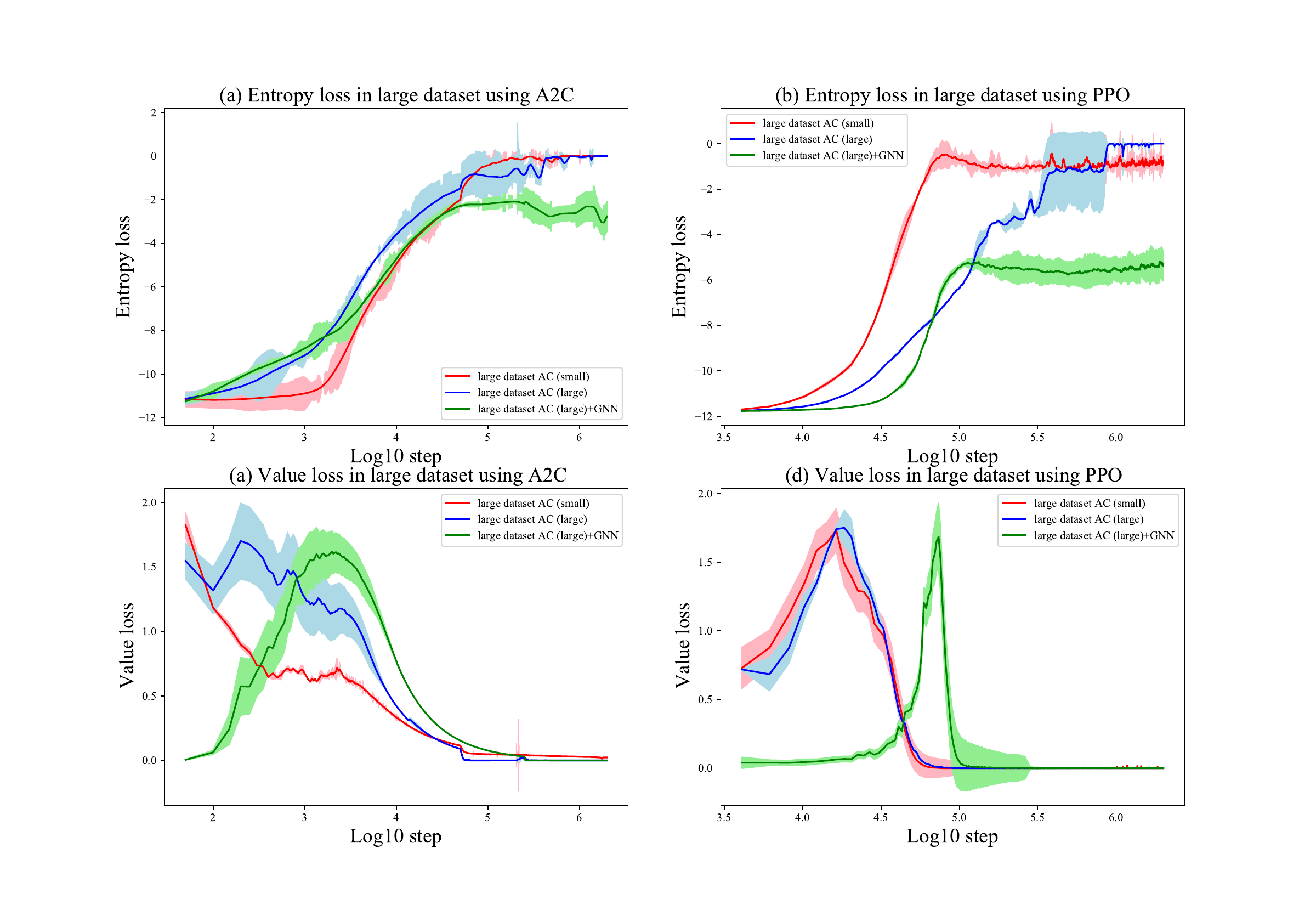}
    \vskip -0.2in
    \caption{{Entropy loss (a,b) and value loss (c,d) of RL training in large dataset using A2C (a,c) and PPO (b,d). The red loss curve captures the agents trained in small action space (AC means action compression). The blue loss curve captures the agents trained in large action space. The green loss curve captures the agents trained in large action space with GNN classifier.}}
    \label{fig:RLmediumdatasetentropyvalueloss}
\end{figure}

Fig. \ref{fig:RLmediumdatasetentropyvalueloss} shows the entropy loss and value loss of training RL agent in a large dataset. 
Without the GNN classifier, it demonstrates convergence in different training settings. 
However, the entropy loss cannot converge to nearly $0$ when training an RL agent with GNN classifier. This is because in the GNN training, we use a threshold to cluster the topologies. Thus, there  can be  multiple topologies with the objective value being larger than the threshold and the final RL policy can sample from multiple  idealized topologies, leading to non-zero entropy loss.  
The value loss still converges to $10^{-7}\sim 10^{-11}$, which confirms the observation.

\vskip -0.2in
\vskip -0.1in

\begin{figure}[H]
    \centering
    \includegraphics[width=0.5\textwidth]{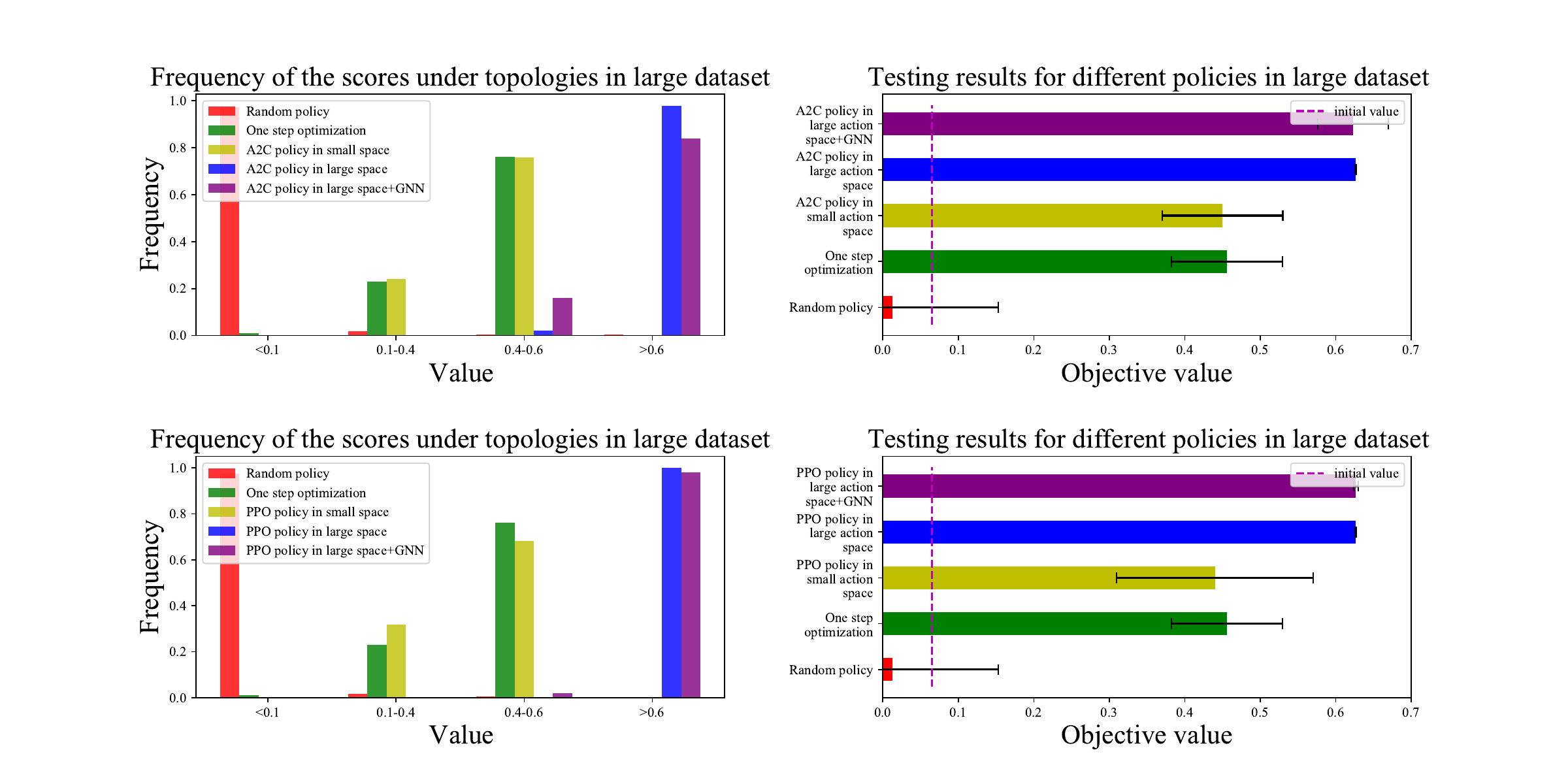}
    \vskip -0.2in
    \caption{{Left: Frequency of the scores under typologies found by different policies in large dataset. 
    Right: Testing results for different policies in large dataset.} }
    \label{fig:mediumdatabar}
\end{figure}

The histogram of the scores in the large dataset is shown in the left figure Fig. \ref{fig:mediumdatabar}. 
The horizontal axis shows the regions of the scores. The vertical axis is the frequency that the score of a chosen topology is in the region.
The  frequencies are computed as follows. Given a random topology, search $30$ topologies by the trained agent, calculate the scores and choose the topology with maximum score. Repeat this process for $5\cdot 10^4$ times  and compute the frequency. 

{We notice that the RL policy (yellow, blue and purple bars) performs significantly better than the random policy (red bar).} The RL policy trained with a small space generates topologies whose scores are between $0.4\sim 0.6$. The RL policy trained in large space generates topologies with scores larger than $0.6$. This means that the RL agent has the ability to find a good topology. 
We also compare the results of the RL agent with one-step optimization. 
In contrast to the small dataset case, where RL and the one-step policy achieve similar performance, here the RL policy performs significantly better than the one-step optimization method. {This illustrates that DRL-GS scales much better than human heuristic methods.} 

\vskip -0.2in
\vskip -0.1in

The right figure in Fig. \ref{fig:mediumdatabar} is the testing results of different methods. The values are calculated in the same way as the right figure in Fig. \ref{fig:frequency_smalldata}.  
The results show that RL policy trained in compressed space gains better average objective values compared with random policy that fails to find an idealized topology in huge full action space.
{Trained in small space, the RL agent gains a similar performance (yellow, A2C: $0.4499\pm 0.0800$, PPO: $0.4399\pm 0.1300$) compared to one step optimization (green, $0.4560\pm 0.0737$).
However, RL agent trained in large space outperforms the performance (blue, A2C: $0.6266\pm 0.0010$, PPO: $0.6266\pm 0.0010$).}
This means that enlarging the searching space helps generate a better policy.  
If the compressed space is chosen appropriately, an RL agent can be trained efficiently and achieve better performance. 

\vskip -0.1in
\begin{figure}[htp]
    \centering
    \includegraphics[width=0.5\textwidth]{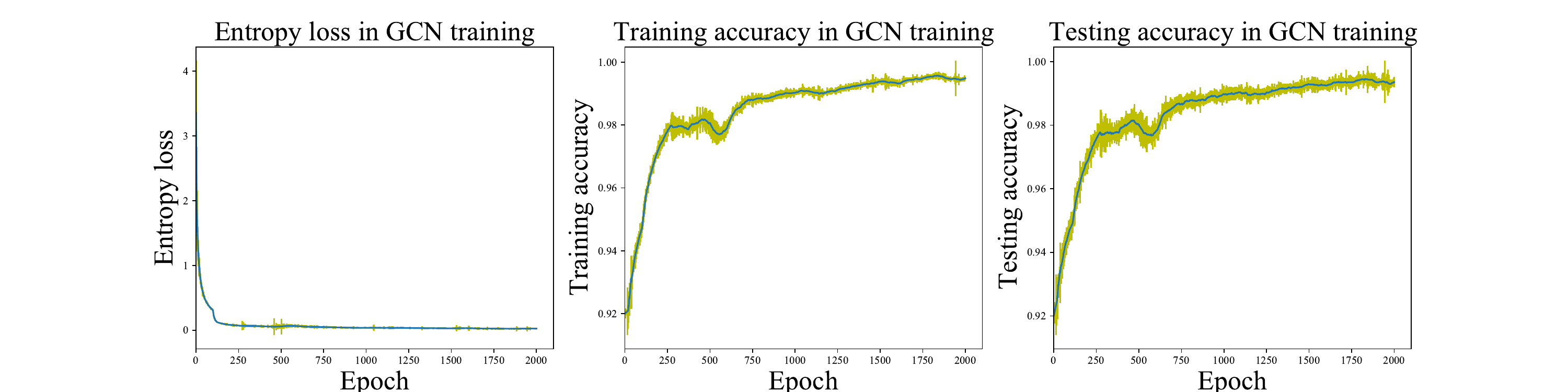}
    \vskip -0.1in
    \caption{GNN training curve with entropy loss, training and testing accuracy in large dataset.}
    \label{fig:gcntrainingmedium}
\end{figure}
\vskip -0.1in

Next we train RL agent in large space using GNN and examine the training and testing accuracy of the GNN classifier.
Different from training GNN in small dataset, we use $q=0.6$ and train the GNN with $2\cdot10^3$ epochs. 
The GNN training results are shown in Fig.  \ref{fig:gcntrainingmedium}.
The entropy loss converges to $10^{-2}$ and the final training accuracy and testing accuracy are higher than $99\%$, which shows that the GNN can classify a topology in high accuracy.

The right figure in Fig. \ref{fig:mediumdatabar} shows the testing results in RL training using the GNN classifier (blue bar). 
Compared to the RL agent trained in large space without GNN, the mean value is slightly smaller and the standard deviation is larger (purple, A2C: $0.6230\pm 0.0471$, PPO: $0.6260\pm 0.030$).
{This is expected, as GNN trades off performance for efficiency when training the DRL algorithm. In addition, PPO achieves superior training performance compared to A2C algorithm because they have different procedures in policy training  \cite{schulman2017proximal}.} 
The advantage of using GNN is that, in cases where  running $\mathtt{Verifier}$ algorithm needs more time, GNN helps improve the efficiency of training RL algorithm, i.e., we need $4$ days to train RL agent with the $\mathtt{Verifier}$ in the large dataset, but we only use $2$ days to train RL agent with the GNN classifier. 
We note that even with a slight performance loss, the RL agent still outperforms the one-step optimization. 

\begin{figure*}[htp]
    \centering
    \includegraphics[width=1.0\textwidth]{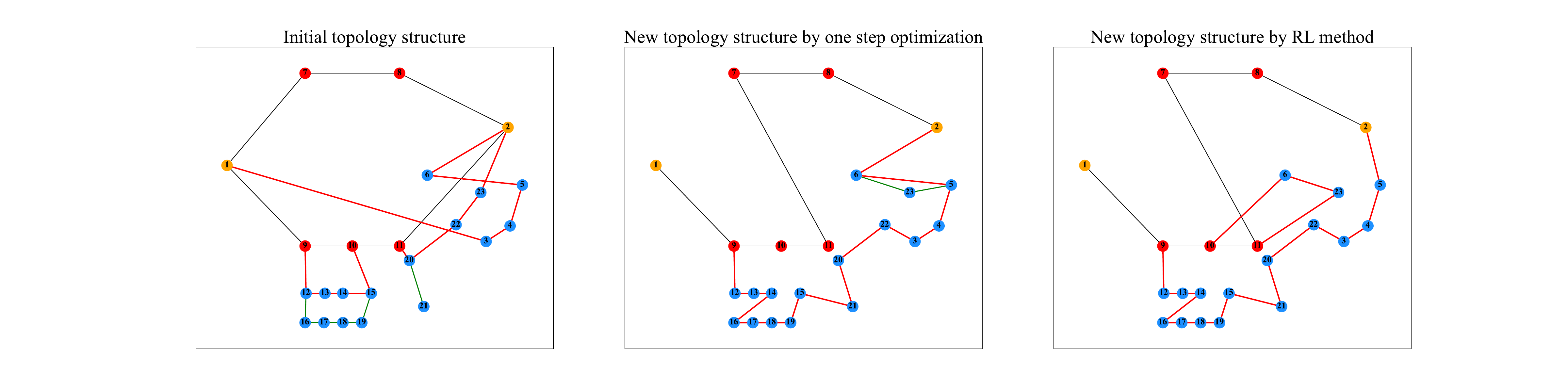}
    \vskip -0.2in
    \caption{The initial topology structure and the optimized topology structure by one-step method and DRL-GS. The type-$T$,type-$H$ and type-$J$ nodes are in orange, red, blue colors, respectively. Initially, there exists a heavily-loaded path  $p_S^{(1)}=\{9\mbox{-}12\mbox{-}13\mbox{-}14\mbox{-}15\mbox{-}10\}$ in red lines attaching nodes $16\sim19$ as elements in the sub path drawn in green lines. 
    At $t=0,Flow(p_S^{(1)},0)=0.12$,
    which is much larger than the value of paths  $p_S^{(2)}=\{1\mbox{-}3\mbox{-}4\mbox{-}5\mbox{-}6\mbox{-}2\}$ ($0.04$) and $p_S^{(3)}=\{2\mbox{-}23\mbox{-}22\mbox{-}20\mbox{-}11\}$ ($0.03$, and $p_S^{(3)}$ includes a hang node $21$). It results in  a low objective value ($0.0653$). 
    One-step optimization generates a new path $p'_S=\{2\mbox{-}6\mbox{-}\cdots\mbox{-}12\mbox{-}9\}$ attaching node $23$ in the sub path with $Flow(p'_S,0)=0.08$. Thus, it achieves a better objective value ($0.4945$). DRL-GS optimizes the topology by generating two new paths $p''^{(1)}_S=\{10\mbox{-}6\mbox{-}23\mbox{-}11\}$,$p''^{(2)}_S=\{2\mbox{-}5\mbox{-}\cdots\mbox{-}12\mbox{-}9\}$  with  $Flow(p''^{(1)}_S,0)=0.072$, $Flow(p''^{(2)}_S,0)=0.08$ achieving a higher objective value ($0.6390$). } 
    \label{fig:optimizationgraph}
\end{figure*}

Fig. \ref{fig:optimizationgraph} shows the topology optimization results in the  $23$-node case.
In the initial topology, there are $2$ primary main paths  $p_P^{(1)}=\{1\mbox{-}7\mbox{-}8\mbox{-}2\}$ and $p_P^{(2)}=\{1\mbox{-}9\mbox{-}10\mbox{-}11\mbox{-}2\}$ and $3$ secondary main paths $p_S^{(1)},p_S^{(2)},p_S^{(3)}$ shown in the figure (we use index $i$ to simplify node $v_i$).
The bandwidth utilization of $p_S^{(1)}$ is large because the maximum utilization of type-$H$ node $9$ and $10$ is low, so $Flow(p_S^{(1)},t)$ is much higher than the benchmark bandwidth utilization $B(t)$ described in \ref{chap:objectivefunction}, and the sub ratio is large because there are $4$ nodes in the sub path. At time $t=0,B(0)=0.07$ while $Flow(p_S^{(1)},0)=0.12$. When $\epsilon=0.4$, we see that $\mathbb{I}\left(\left|\frac{Flow(p_S^{(1)},0)}{B(0)}-1\right|\le\epsilon\right)=0$ and $Sratio(p_S^{(1)})=0.4$, 
which suffers a loss. This unbalance among paths results in a low objective value ($0.0653$). One-step optimization constructs two paths, 
a primary main path $p'_{P}=\{1\mbox{-}9\mbox{-}\cdots\mbox{-}8\mbox{-}2\}$ and a secondary main path $p'_S$ shown in the figure. 
The maximum utilization of node $2$ is larger than that of node $9$, so this path can bear traffic pressure   $Flow(p'_S,0)=0.08$, %
which is close to $B(0)$. 
Hence, it achieves a $0.4945$ objective value.
After training the RL agent, 
RL policy finds that generating two paths
$p''^{(1)}_S,p''^{(2)}_S$ (shown in the figure) helps because new paths guarantee that both of the bandwidth utilization are close to $B(0)$, i.e.,  $Flow(p''^{(1)}_S,0)=0.072,Flow(p''^{(2)},0)=0.8$. So it achieves a higher objective value $0.6390$.

\section{Conclusion}

In this paper, we consider the problem of network topology optimization with management constraints. 
We propose a deep reinforcement learning framework called DRL-GS, which consists of a verification model to testify the topology, calculate the objective value and generate data, an action compression method to eliminate searching space, and a GNN classifier to replace the $\mathtt{Verifier}$ algorithm to enable efficient topology search. %
We conduct extensive experiments based on small scale and large scale datasets. 
The results show that DRL-GS outperform the human-expert based one-step optimization method in finding optimal network topologies.

\section*{Acknowledgments}
The work of Zhuoran Li, Ling Pan and Longbo Huang is supported by the Technology and Innovation Major Project of the Ministry of Science and Technology of China under Grant 2020AAA0108400 and 2020AAA0108403, the Tsinghua University Initiative Scientific Research Program, and Tsinghua Precision Medicine Foundation 10001020109.
We thank China Mobile Research Institute for technical support, including data information and the Jiutian Artificial Intelligence Platform for experiments.

\bibliographystyle{unsrt}
\bibliography{IEEEabrv}

\section{Appendix: One-step optimization pesudo codes}
Here we propose one-step optimization method by constructing new components as the benchmark algorithm to compare with DRL-GS. 
It is useful in a small dataset while in a large topology network, 
it would suffer failure. The details of this algorithm are described as algorithm \ref{alg:onestep}.

\begin{algorithm*}[htp]
\begin{algorithmic}[1]
\caption{One step optimization method}
\label{alg:onestep}
\STATE \textbf{Input}: Topology $G=(V,E)$, benchmark $B(t)$, threshold value $D_{\max},\epsilon$, horizon $T=24$.
\FOR{$Type=H,J$}
\STATE Generate $G'=(V',E')$ that $V'=\{v|v\in V,Type(v)=Type\},E'=\{e|e=(v_i,v_j),Dist(e)\le D_{max},v_i,v_j\in V'\}$. 
\STATE Run $\mathtt{BFS}$ algorithm in $G'$ to generate buffer  $\mathcal{D}_{Type}=\{D_i=\{v_{i_1},...,v_{i_n}\}\}_{i=1}^{d}$. 
\WHILE{$\exists D_i, |D_i|>0$}
\STATE For every group $D_i$, sort nodes based on $\sum_{t}Flow(v_i,t)$ in decrease order as $v_{i_1}',...,v_{i_n}'$. For every node group $v_{i_1}',...,v_{i_u}'$, where $u=1,...,n$, calculate: $$ Utility(u)=\frac{1}{T}\sum_{t=1}^{T}\mathbb{I}\left(\left|\frac{\sum_{v_k=v_{i_1}',...,v_{i_u}'}Flow(v_k,t)}{\max_{v\in V}U_{\max}(v)}-B(t)\right|/B(t)\le\epsilon\right) $$

\STATE Set that $U=\text{argmax}_{u} Utility(u),u_0=\max_{u}(U)$, $D_i\leftarrow D_i\backslash\{v_{i_1}',...,v_{i_{u_0}}'\}$. For group $D'' = \{v_{i_1}',...,v_{i_{u_0}}'\}$, run $\mathtt{BFS}$ algorithm in $G''=(D'',E''),(\forall e\in E'',Dist(e)\le D_{\max})$ to search the connected component $\tilde{D}_1,...,\tilde{D}_{\tilde{d}}$ and add these into the buffer $\mathcal{D}_{Type}$.
\ENDWHILE
\ENDFOR
\STATE Set edge buffer $\mathcal{E}$. $\forall D_h\in \mathcal{D}_H, D_j\in\mathcal{D}_J$, generate $G_h=(D_h,E_h),G_j=(D_j,E_j)$ for connectivity. If $\exists v_{h_{1}},v_{h_{2}}\in D_h, v_{j_{1}},v_{j_{2}}\in D_j$ that $Dist(e_i)\le D_{\max},e_i=(v_{j_{i}},v_{h_{i}}),i=1,2$, add $e_i$ into $\mathcal{E}$. Generate $\tilde{G}=(V,\tilde{E})$ that $\tilde{E}=(\cup_{h}E_h)\cup(\cup_{j}E_j)\cup\mathcal{E},\forall E_h,E_j$.
\STATE \textbf{Return} $\tilde{G}$.
\end{algorithmic}
\end{algorithm*}




 

\vspace{-20pt}
\begin{IEEEbiography}[{\includegraphics[width=1in,height=1.25in,clip,keepaspectratio]{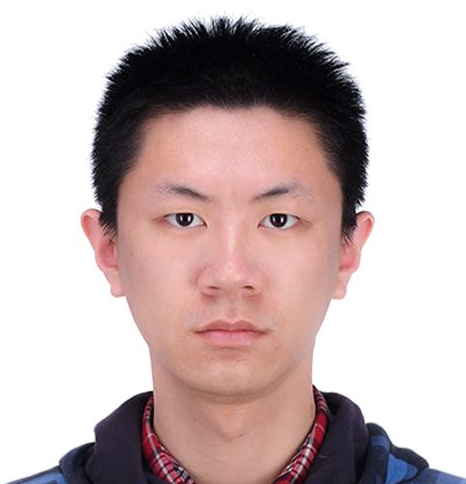}}]{Zhuoran Li}
received the B.E. degree in 2020
in electronic engineering from Tsinghua University,
Beijing, China, where he is currently working toward
the Ph.D. degree in the Institute for Interdisciplinary Information Sciences (IIIS) (headed by Prof. Andrew Yao), Tsinghua University, advised by Prof. Longbo Huang.
\end{IEEEbiography}
\vspace{-30pt}
\begin{IEEEbiography}[{\includegraphics[width=1in,height=1.25in,clip,keepaspectratio]{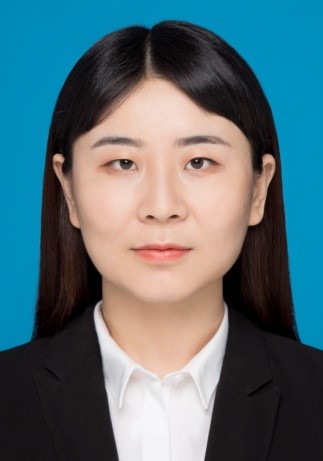}}]{Xing Wang}
received the MS degree from the Beijing Normal University, Beijing, China, in 2018. She is currently a technical researcher at the Artificial Intelligence and Intelligent Operation Center of China Mobile Research Institute. Her research interests include network intelligence, network topology optimization, spatial-temporal data mining, reinforcement learning, and deep learning.
\end{IEEEbiography}
\vspace{-30pt}
\begin{IEEEbiography}[{\includegraphics[width=1in,height=1.25in,clip,keepaspectratio]{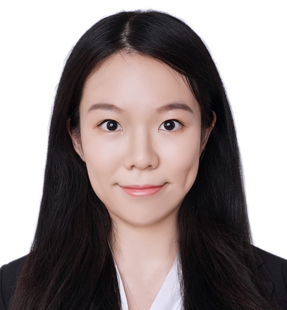}}]{Ling Pan}
is a postdoctoral fellow at MILA supervised by Prof. Yoshua Bengio. She received her Ph.D. from the Institute for Interdisciplinary Information Sciences (IIIS), Tsinghua University in 2022, advised by Prof. Longbo Huang. She also spent time at Stanford University, University of Oxford, and the Machine Learning group at Microsoft Research Asis during her Ph.D. study. Her research interests include generative flow networks (GFlowNets), deep reinforcement learning, and multi-agent systems.
\end{IEEEbiography}
\vspace{-30pt}
\begin{IEEEbiography}[{\includegraphics[width=1in,height=1.25in,clip,keepaspectratio]{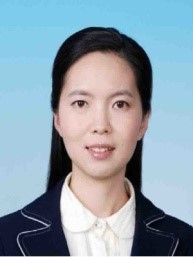}}]{Lin Zhu}
received the Ph.D. degree from the Beijing Institute of Technology (BIT), Beijing, China in 2008. She is currently senior technical researcher in artificial intelligence algorithm of China Mobile Research Institute. Her research interests include intelligent network, intelligent pipeline and communication network. She has applied for more than 30 national invention patents in related filed.
\end{IEEEbiography}
\vspace{-30pt}
\begin{IEEEbiography}[{\includegraphics[width=1in,height=1.25in,clip,keepaspectratio]{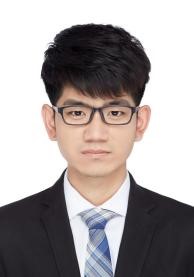}}]{Zhendong Wang}
received the M.S. degree and the Ph.D. degree from Beijing Institute of Technology, Beijing, China, in 2016 and 2021 respectively. He is currently an algorithm engineer at Artificial Intelligence and Intelligent Operation Center of China Mobile Research Institute. His research interests include artificial intelligence, spatio-temporal data modeling.
\end{IEEEbiography}
\vspace{-30pt}
\begin{IEEEbiography}[{\includegraphics[width=1in,height=1.25in,clip,keepaspectratio]{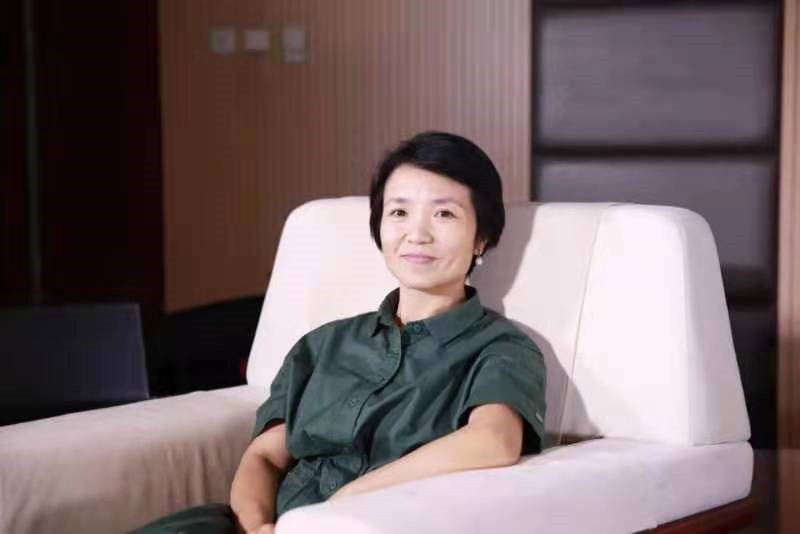}}]{Junlan Feng}
is Vice Chairman of the China Artificial Intelligence Industry Alliance, Chief Scientist at China Mobile, General Manager of AI and Intelligent Operation R\&D Center, Board Chair of Linux Foundation Network.
Dr. Feng received her Ph.D. on Speech Recognition from Chinese Academy of Sciences, and joined AT\&T Labs Research in 2001, as a principal researcher on Speech recognition, language understanding and data mining until 2013. Dr. Feng has led R\&D of China Mobile on artificial intelligence and intelligent operation since September 2013. 
She is an IEEE senior member. She had served as an IEEE speech and language committee member and an IEEE industry committee member. She is a frequent reviewer and organizer for major data mining, speech, and natural language international conferences and journals. Dr. Feng has over 100 professional publications and co-authored a book. She holds 51 issued U.S and international patents and 3 issued Chinese patents. She has 46 pending patent applications.
Dr. Feng has won more than 20 domestic and foreign R\&D awards, including AT\&T CTO Award in 2009. “JiuTian”, the AI platform developed by her team won the single product gold award of China Mobile in 2019, the second prize of scientific and technological achievements of China Institute of Communications, the third prize of scientific and technological progress of China Institute of Electronics, and the 2018 Innovation Project Award of Deep Integration of Artificial Intelligence and Real Economy by the Ministry of Industry and Information Technology.
\end{IEEEbiography}
\begin{IEEEbiography}[{\includegraphics[width=1in,height=1.25in,clip,keepaspectratio]{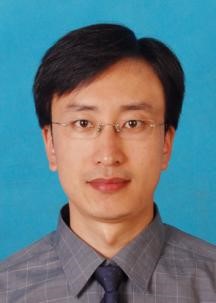}}]{Chao Deng}
received the M.S. degree and the Ph.D. degree from Harbin Institute of Technology, Harbin, China, in 2003 and 2009 respectively. He is currently a deputy general manager at the Artificial Intelligence and Intelligent Operation Center of China Mobile Research Institute. His research interests include machine learning and artificial intelligence for ICT operations.
\end{IEEEbiography}
\vspace{-30pt}
\begin{IEEEbiography}[{\includegraphics[width=1in,height=1.25in,clip,keepaspectratio]{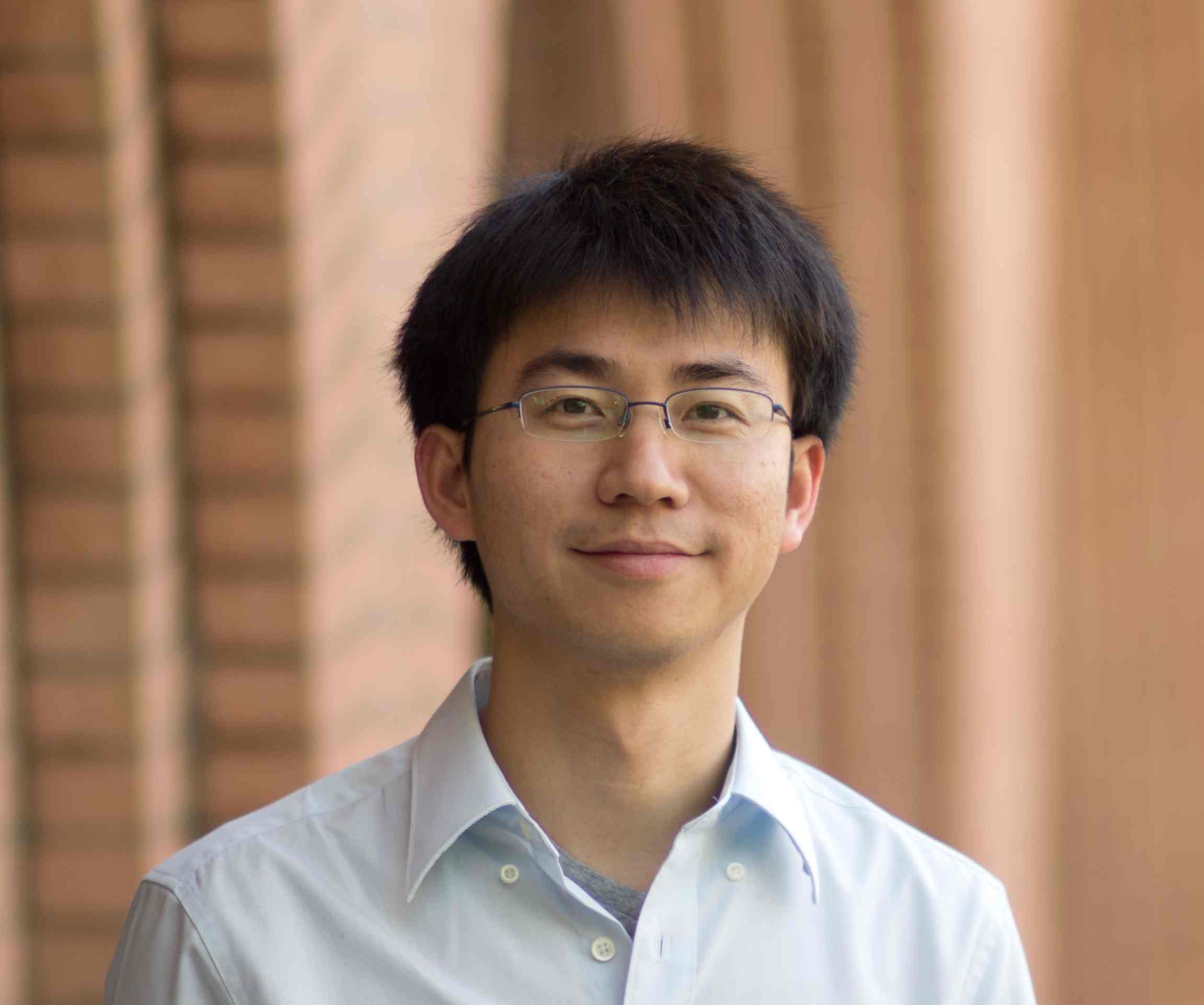}}]{Longbo Huang}
Dr. Longbo Huang is an associate professor (with tenure) at the Institute for Interdisciplinary Information Sciences (IIIS) at Tsinghua University, Beijing, China. His research focuses on decision intelligence (AI for Decisions). Dr. Huang has held visiting positions at the LIDS lab at MIT, the Chinese University of Hong Kong, Bell-labs France, and Microsoft Research Asia (MSRA). He was a visiting scientist at the Simons Institute for the Theory of Computing at UC Berkeley in Fall 2016. Dr. Huang serves/served on the editorial board for IEEE JSAC Special Issue on Human-in-the-loop Mobile Network (Lead guest editor 2016), IEEE TCOM (2017-2020), ACM ToMPECS (2017-present), Elsevier PEVA (2022-present) and IEEE/ACM ToN (2019-present). Dr. Huang is an ACM Distinguished Scientist, CCF Distinguished Member, IEEE Senior Member, an IEEE ComSoc Distinguished Lecturer and an ACM Distinguished Speaker. Dr. Huang received the Outstanding Teaching Award from Tsinghua university in 2014. He received the Google Research Award and the Microsoft Research Asia Collaborative Research Award in 2014, and was selected into the MSRA StarTrack Program in 2015. Dr. Huang won the ACM SIGMETRICS Rising Star Research Award in 2018.
\end{IEEEbiography}


\vfill

\end{document}